%% file: RS_arxiv.tex
\documentclass[12pt]{article}
\usepackage{amsmath,amssymb,amsthm}
\usepackage{graphicx,epsf, color}
\usepackage{enumerate}
\usepackage[sort]{natbib}
\usepackage{url} 
\usepackage{multirow}

\usepackage{times}
\usepackage{bm}

\usepackage[plain,noend]{algorithm2e}

\makeatletter
\renewcommand{\algocf@captiontext}[2]{#1\algocf@typo. \AlCapFnt{}#2} 
\def\@algocf@capt@plain{top}
\renewcommand{\algocf@makecaption}[2]{%
  \addtolength{\hsize}{\algomargin}%
  \sbox\@tempboxa{\algocf@captiontext{#1}{#2}}%
  \ifdim\wd\@tempboxa >\hsize
    \hskip .5\algomargin%
    \parbox[t]{\hsize}{\algocf@captiontext{#1}{#2}}
  \else%
    \global\@minipagefalse%
    \hbox to\hsize{\box\@tempboxa}
  \fi%
  \addtolength{\hsize}{-\algomargin}%
}
\makeatother

\usepackage{subfig}

\input{Notation}

\def\mR{\mathbb R}

\newtheorem{assumption}{Assumption}
\newtheorem{definition}{Definition}[section]

\newtheorem{theorem}{Theorem}
\newtheorem{lemma}{Lemma}
\newtheorem{corollary}{Corollary}

\newcommand\NL{{\it Node $\ell_1$}}
\newcommand\CL{{\it Child $\ell_2$}}
\newcommand\DL{{\it Descendant $\ell_2$}}
\newcommand\SPG{{smoothing proximal gradient}}
\newcommand\Prob{{pr}}
\newcommand{\st}[1]{} 
\newcommand{\mathst}[1]{} 
\def\red{\color{black}}

\usepackage{authblk}

\begin{document}

\def\spacingset#1{\renewcommand{\baselinestretch}%
{#1}\small\normalsize} \spacingset{1}

\pagenumbering{arabic}
\title{\bf
It's All Relative: New Regression Paradigm for Microbiome Compositional Data}

\author[1]{Gen Li\thanks{Corresponding Author: ligen@umich.edu}}
\author[2]{Yan Li}
\author[2]{Kun Chen}

\affil[1]{Department of Biostatistics, School of Public Health, University of Michigan, Ann Arbor}
\affil[2]{Department of Statistics, University of Connecticut}
\renewcommand\Authands{ and }

  \maketitle

\bigskip

\begin{abstract}
Microbiome data are complex in nature, involving high dimensionality, compositionality, zero inflation, and taxonomic hierarchy.
Compositional data reside in a simplex that does not admit the standard Euclidean geometry.
Most existing compositional regression methods rely on transformations that are {inadequate or even inappropriate} in modeling data with excessive zeros and taxonomic structure. We develop a novel relative-shift regression framework  that directly uses compositions as predictors.
The new framework provides a paradigm shift for compositional regression and offers superior biological interpretation.
New equi-sparsity and taxonomy-guided regularization methods and an efficient smoothing proximal gradient algorithm are developed to facilitate feature aggregation and dimension reduction in regression.
As a result, the framework can automatically identify clinically relevant microbes even if they are important at different taxonomic levels.
A unified finite-sample prediction error bound is developed for the proposed regularized estimators.
We demonstrate the efficacy of the proposed methods in extensive simulation studies.
The application to a preterm infant study reveals novel insights of association between the gut microbiome and neurodevelopment.
\end{abstract}

\noindent%
\vfill

\newpage
\spacingset{1.45} 


\section{Introduction}

The human microbiome plays a critical role in human health and disease \citep{nih2019review}.
Modern technology enables cost-effective acquisition of microbiome data.
Large collaborative efforts such as the Human Microbiome Project\st{ (HMP)} \citep{turnbaugh2007human} have generated rich and valuable databases.
The ever-increasing microbiome data allow us to better decipher the relationship between the microbiome and host health.
A thorough understanding of the link between the microbiome and health outcomes promises to revolutionize diagnosis and prognosis and may lead to therapeutic breakthroughs.

However, microbiome data are complex in nature, {involving high dimensionality, compositionality, zero inflation, and taxonomic hierarchy.}
Microbiome data are usually measured through high-throughput sequencing (e.g., 16S rRNA sequencing).
High dimensional read counts are then clustered into hundreds of ``operational taxonomic units'' (OTUs) for subsequent analyses.
The corresponding OTU table provides the abundance of each taxon (i.e., feature) in each community (i.e., sample).
Due to the large variation in library size (i.e., total number of reads in each sample), read counts in OTU table are usually normalized as compositions (i.e., relative abundance) \citep{gloor2016s,tsilimigras2016compositional}.
Compositionality is a prominent feature of microbiome data that complicates statistical analyses.
As compositions reside in a simplex that does not admit the standard Euclidean geometry, many standard notions and methods do not directly apply.
In addition, microbiome data are usually highly sparse with few dominant compositions and excessive zeros \citep{xia2018modeling,xu2020zero}.
There also exists extrinsic information such as the taxonomic hierarchy among taxa.
The taxonomic tree captures the classification of microbes at different ranks.
OTU data at lower taxonomic ranks have higher resolution (i.e., more features) but are more prone to measurement errors, while data at higher taxonomic ranks have lower resolution with higher accuracy.
Namely, there is a trade-off between data resolution and accuracy along the taxonomic hierarchy, which provides useful guidance for microbiome analysis.


{The unique features of microbiome data pose new challenges for statistical analysis \citep{li2015microbiome,zhou2019review}. Here we focus on regression analysis with microbiome compositional data, which is critical to study the association of bacterial taxa and clinical outcomes.}
Due to the compositional nature, microbiome data are typically transformed (e.g., log-ratio transformation) first to admit the Euclidean geometry \citep{aitchison2005compositional}.
Linear regression models are subsequently built upon transformed data.
For example, one of the most commonly used models is the log-contrast model \citep{aitchison1984log} where log-ratio-transformed compositions are used as predictors in a linear regression.
An equivalent symmetric form of the model is as follows
\[
y=\beta_0+\beta_1\log{x_1}+\cdots+\beta_p\log{x_p}+\varepsilon,
\]
where the compositional vector $\bx$ is in the $(p-1)$-simplex $\simp^{p-1}=\{\bx{\red=(x_1, \ldots, x_p)^T}\in\real^p: \sum_{j=1}^p x_j=1, x_j\geq0, j=1,\ldots,p\}$ and the coefficients satisfy a linear constraint $\sum_{j=1}^{p}\beta_j=0$.
The model enjoys the subcompositional coherence and scale and permutation invariance properties \citep{aitchison1982statistical}.
\cite{lin2014variable} and \cite{shi2016regression} proposed variable selection methods for such models.
\cite{lu2019generalized} further generalized it to non-Gaussian responses.
Centered log-ratio transformation is also frequently used in the literature and it has been shown to be equivalent to the log-contrast model if the zero-sum constraint is imposed on the regression coefficients \citep{randolph2018kernel,wang2017structured}. 

Although almost all existing compositional regression methods are based on transformed data, the transformation procedure itself has several major drawbacks.
Most prominently, the commonly used logarithmic transformation cannot handle zero values.
A common practice is to artificially replace zero with some preset small value to avoid singularity \citep{aitchison1984log,Palarea2013, lin2014variable}.
However, since microbiome data are typically inflated with zeros, such manipulation may introduce unwanted bias and result in misleading results \citep{mcmurdie2014waste}.
Another drawback is the lack of straightforward biological interpretation.
The log transformation removes compositions from the simplex, but does not eliminate the interrelated structure of the data.
The change in one predictor value is linked with the change in at least one other predictor value.
As a result, one cannot simply interpret the coefficient $\beta_j$ as the effect size corresponding to one unit increase in $\log x_j$ with others held fixed. 

In addition, the transformation hinders the incorporation of extrinsic information such as the taxonomic tree structure.
Several attempts have been made in the literature, but there is no consensus about how to properly regularize the regression coefficients to reflect extrinsic information.
This is largely due to the lack of clear interpretation of the coefficients for transformed data.
For example, \cite{shi2016regression} proposed a subcompositional model that partially accounts for the taxonomic structure. However, it can only handle two taxonomic ranks.
\cite{garcia2013identification} and \cite{wang2017structured} developed group-lasso-type regularization methods to achieve subcomposition selection.
\cite{randolph2018kernel} proposed to translate extrinsic information into kernels and incorporate them into a penalized regression framework.
However, kernelizing taxonomy or phylogeny may oversimplify the data structure since the original tree structure cannot be fully characterized by a similarity matrix.
Besides, to the best of our knowledge, no existing method ensures compatible results across taxonomic ranks.
That is, analyses conducted on the same OTU data at different taxonomic ranks may have drastically different results.
For example, a species may be deemed important from the species-level analysis, but the genus it belongs to may have negligible effect from the genus-level analysis.
Such discrepancy may call microbiome regression analysis into question.

In this paper, we break new ground to develop a new regression paradigm for microbiome compositional data.
The new framework, called {\em Relative-Shift}, directly models compositions as predictors without transformation.
It is fundamentally different from the log-contrast models and provides an alternative approach to microbiome regression.
The basic model is based on a simple yet intriguing finding, that is, the regression on compositions is completely identifiable if we just eliminate the intercept term. 
Namely, an intercept-free linear regression model with compositional predictors is the basic form of our proposed relative-shift model.
Although seemingly simple, the model carries important biological interpretations and enjoys many desirable properties such as scale and shift invariance.
In particular, the contrast in regression coefficients can be interpreted as the effect size of shifting concentrations between taxa (i.e., the origination of the name, {\em relative-shift}).

The relative-shift model also serves as a flexible basis for accommodating unique features of microbiome data.
For example, zero values are directly handled without substitution; {high dimensional compositional features can be reduced through aggregation or amalgamation.}
More importantly, the taxonomic tree structure among taxa can be tactfully incorporated as well.
In particular, we develop new taxonomy-guided regularization methods for parameter estimation.
The proposed methods can adaptively determine what taxa at which taxonomic levels are
most relevant to the response.
They are robust against the change in the taxonomic levels of the study and offer superior biological interpretability.
The methods utilize data across all taxonomic ranks and strike a good balance between data resolution and accuracy.

\st{
The rest of the paper is organized as follows. In Section {\ref{sec:RS}}, we introduce the relative-shift model.  In Section {\ref{sec:reg}}, we develop novel regularization methods for feature aggregation in high dimension with and without taxonomic guidance. In Section {\ref{sec:comp}}, we devise model fitting algorithms for different settings.
In Section {\ref{sec:theory}}, we derive a unified finite-sample prediction error bound for the proposed estimators.
Comprehensive simulation studies are contained in Section {\ref{sec:sim}} and a real data application to a microbiome study of neurodevelopment in preterm infants is in Section {\ref{sec:real}}.
We conclude and discuss open questions in Section {\ref{sec:dis}}.
}

\section{Relative-Shift Regression Paradigm}\label{sec:RS}

Let $\by = {\red(y_1,\ldots,y_n)^T} \in\real^{n}$ denote the continuous response vector of $n$ samples.
Let $\bx_i=(x_{i1},\ldots,x_{ip})^T\in\simp^{p-1}$ represent the microbial compositional vector of $p$ taxa and  $\bc_i=(c_{i1},\ldots, c_{iq})^T\in\real^q$ be a length-$q$ auxiliary non-compositional covariate vector for the $i$th subject ($i=1,\ldots,n)$.

We propose the following relative-shift model for compositional regression with covariate adjustment
\be\label{RS}
y_i=\bc_i^T\bbeta_c  + \beta_1 x_{i1}+\cdots+\beta_p x_{ip}+\varepsilon_i,
\ee
where $\varepsilon_i$ is the random noise with mean zero and variance $\sigma^2$, and $\bbeta_c\in\real^q$ and $\bbeta=(\beta_1,\ldots,\beta_p)^T\in\real^p$ are coefficient vectors for covariates and compositions, respectively.
The relative-shift model is identical to a linear regression model less the intercept term, yet the difference ensures the identifiability of the model.

In contrast to the models based on transformed data, the relative-shift model uses proportions as predictors, and thus directly characterizes how composition changes affect the response. 
The coefficients for compositions provide important biological interpretation.
We stress that they shall not be interpreted separately since compositions are interrelated.
Instead, differences between coefficients are readily interpretable.
For example, for any pair of taxa $(j,k)$, we can write $\beta_jx_{\cdot j} + \beta_kx_{\cdot k} = \beta_k(x_{\cdot j} + x_{\cdot k}) + (\beta_{j} - \beta_{k})x_{\cdot j}$. Therefore, $(\beta_{j} - \beta_{k})$ can be interpreted as the effect size of $x_{\cdot j}$ on the outcome when holding $x_{\cdot j} + x_{\cdot k}$ fixed, that is, $\Delta(\beta_j-\beta_k)$ can be viewed as the effect on the response with a shift of $\Delta$ concentration from the $k$th taxon to the $j$th taxon 
while holding other taxa fixed.  As another example, consider the triplet of taxa $(j,k, r)$. Write $\beta_jx_{\cdot j} + \beta_rx_{\cdot r} + \beta_jx_{\cdot k} = \beta_r(x_{\cdot j} + x_{\cdot r}+x_{\cdot k}) + (\beta_j -\beta_r)x_{\cdot j} + (\beta_k-\beta_r)x_{\cdot k}$. Then a contrast of the form $\Delta(a\beta_j + (1-a)\beta_k - \beta_r)$ can be interpreted as the effect on the response with a shift of $\Delta$ concentration from the $r$th taxon to the $j$th and $k$th taxa by the amount $a\Delta$ and $(1-a)\Delta$ $(0\leq a\leq 1)$, respectively, while holding the other taxa fixed.
Intriguingly, under the proposed model, all the contrasts of the regression coefficients can be interpreted as the effect of certain shifts of abundances of a group of taxa while holding their sum fixed. This is the origination of the name \textit{relative-shift regression}.

Although simple, the relative-shift model well characterizes the fundamental relations between compositional predictors and the response. Surprisingly, this simple alternative has been overlooked in the microbiome literature.
The relative-shift model enjoys several desirable properties. First, it is scale and shift invariant. In particular, if the response is multiplied by a constant, the model will remain the same if all coefficients are multiplied by the same constant. If the response  
shifts by a constant, the effect can be offset by adding the same constant to all coefficients for the compositions. This invariance also indicates that the magnitude or the absolute value of the coefficients is not important. Instead, the relative relationships between different parameters are crucial.
Second, the model is immune to zero inflation. It can directly handle zeros in the design matrix without taking additional steps as in a log-contrast model. 
Third, equal coefficients directly translate into feature aggregation. Namely, if the coefficients for two (or more) taxa are the same, they can be directly combined as a new entity with the new composition being the sum of the original compositions and the coefficient being the shared coefficient. This serves as the foundation for the regularization methods in Section \ref{sec:reg}. 


\section{Regularization Methods for Feature Aggregation}\label{sec:reg}
\subsection{Equi-Sparsity Regularization}\label{subsec:hd}
To estimate model parameters, one can directly resort to the ordinary least squares approach by minimizing the following convex objective function
\[
g(\bbeta_c,\bbeta)={1\over 2n}\|\by-\bC\bbeta_c-\bX\bbeta\|^2,
\]
where $\bC=(\bc_1,\ldots,\bc_n)^T\in\real^{n\times q}$ is a covariate matrix, $\bX=(\bx_1,\ldots,\bx_n)^T{\red\in\real^{n\times p}}$ is a compositional matrix, and $\|\cdot\|$ represents the Frobenius norm.
The optimization has a unique closed-form solution if there is no collinearity in $\bC$ and $\bX$.
However, microbiome data are often high dimensional, with the number of taxa $p$ greater than the sample size $n$.
As a result, the design matrix becomes singular and there is no unique solution to the problem.

To address the problem, we introduce an {\em equi-sparsity} regularization for { compositional coefficients}.
Equi-sparsity is a generalization of the widely studied zero-sparsity \citep{she2010sparse}.
It encourages coefficients to be equal to each other (i.e., clustering of coefficients) rather than close to zero.
In the relative-shift Model \eqref{RS}, equi-sparsity for composition coefficients is especially relevant because we only care about the relative relations between coefficients rather than their absolute numerical values.
If two coefficients are equal, shifting concentrations between the corresponding pair of taxa does not change the model.
Correspondingly, the two taxa can be combined to form a new entity without losing any information.
For example, if $\beta_j=\beta_k$, the corresponding taxa $j$ and $k$ can be directly combined since $\beta_jx_{ij}+\beta_kx_{ik}=\beta_j(x_{ij}+x_{ik})$.
As a result, equi-sparsity achieves dimension reduction of compositional data by feature aggregation.

More specifically, we consider the following equi-sparsity regularization for the composition coefficients $\bbeta$ in Model \eqref{RS}
\bes
\pen_E(\bbeta)=\sum_{j<k}\omega_{jk}|\beta_j-\beta_k|,
\ees
where $\omega_{jk}$ is some predefined positive weight between taxa $j$ and $k$.
The weights can be determined based on extrinsic information (e.g., phylogenetic information), where a larger value induces more penalty on the pairwise difference and vice versa.
By default, we set all weights to be equal to 1, and the penalty term reduces to the clustered lasso penalty studied in  \cite{she2010sparse}.
It also coincides with the graph-guided-fused-lasso penalty in \cite{kim2009multivariate} with a complete graph.

Applying the equi-sparsity regularization to Model \eqref{RS}, we obtain the following convex optimization problem for parameter estimation in high dimension
\be\label{opt1}
(\widehat{\bbeta_c},\widehat{\bbeta})=\argmin_{\bbeta_c,\,\bbeta}\quad g(\bbeta_c,\bbeta)+\lambda\pen_E(\bbeta),
\ee
where $\lambda>0$ is a tuning parameter.
We  devise an efficient algorithm for solving \eqref{opt1} in Section \ref{sec:comp}.

\subsection{Taxonomic-Tree-Guided Regularization}\label{subsec:tax}
Assume the taxonomic tree is available as prior knowledge.
Now we elaborate how to incorporate the taxonomic tree structure into the relative-shift regression paradigm via novel regularization methods.
The basic idea of the taxonomic-tree-guided regularization is to encourage equi-sparse coefficients for taxa sharing similar taxonomic paths.
In other words, the more common ancestors two taxa share, the more likely they are to be aggregated.

Let $T$ represent a $p$-leafed taxonomic tree, $I(T)$ represent the set of internal nodes, $L(T)$ represent the set of leaf nodes, and $|T|$ represent the total number of nodes in a tree.
We follow the commonly used notions of child, parent, sibling, descendant and ancestor to describe the relations between nodes.
Each leaf node of the tree corresponds to a taxon and each internal node corresponds to a group of taxa (i.e., the descendant leaf nodes of the internal node).

\begin{figure}[hbpt]
\centering
       \includegraphics[width=2.5in]{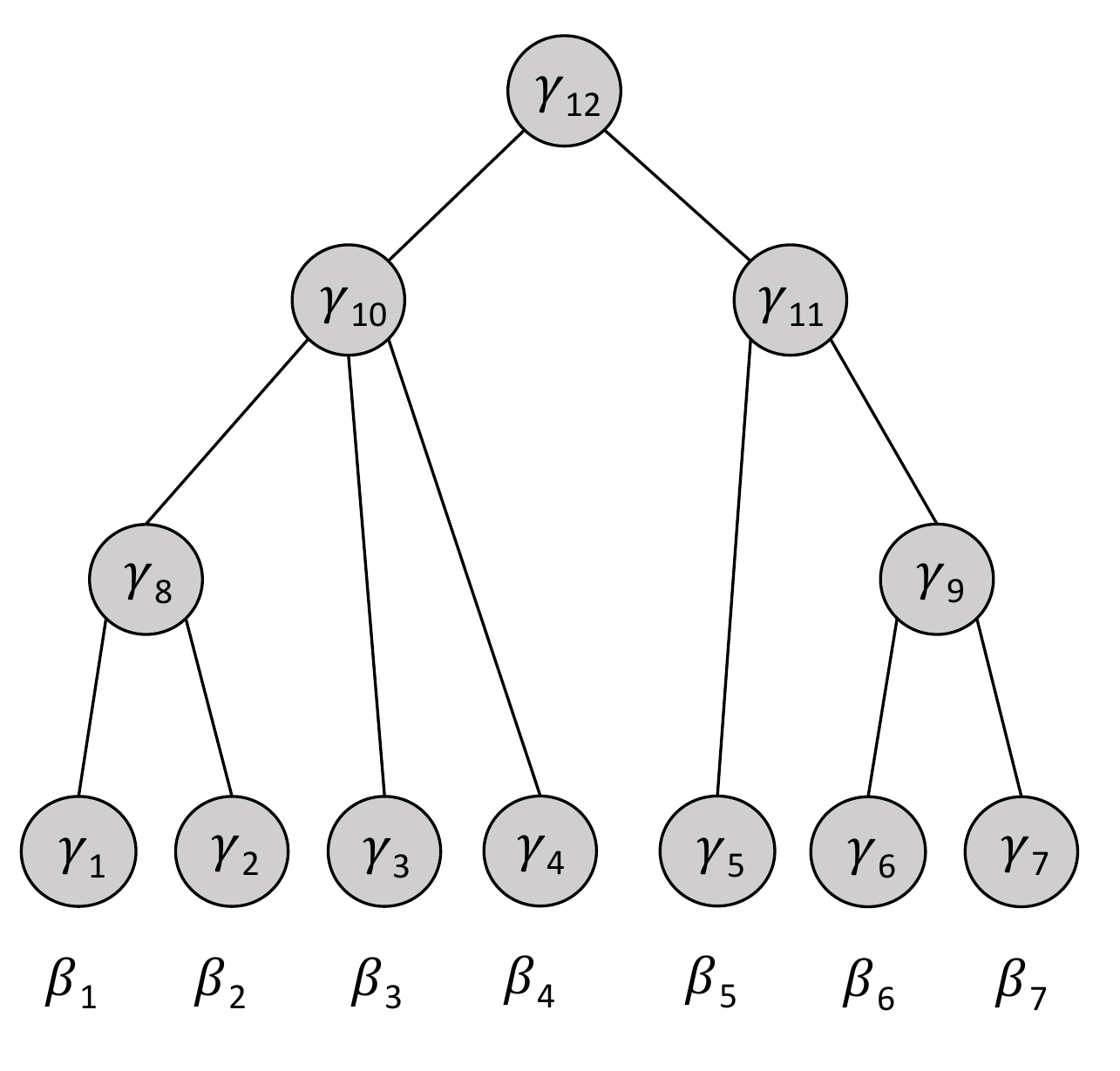}
\caption{Illustration of tree-guided reparameterization.}
\label{fig:tree}
\end{figure}

\cite{yan2018rare} recently proposed a tree-guided regularization method  for rare feature aggregation. 
A significant innovation is the adoption of a tree-based parameterization where original regression coefficients are broken down into intermediate coefficients assigned to each node of a tree.
For example, Figure \ref{fig:tree} provides an illustration of a tree $T$ with seven leaf nodes (one for each regression coefficient).
An intermediate coefficient $\gamma_u$ is assigned to each node $u\in T$.
Then each original coefficient $\beta_j$ (at the leaf node $j$) is expressed as
\[
\beta_j=\sum_{u\in \mbox{\footnotesize Ancestor}(j)\cup \{j\}} \gamma_u,
\]
where $\mbox{Ancestor}(j)$ denotes the set of ancestors of node $j$.
For example, $\beta_1=\gamma_1+\gamma_8+\gamma_{10}+\gamma_{12}$.
Correspondingly, the coefficient vector $\bbeta$ can be written as a linear transformation of the intermediate coefficient vector $\bgamma=(\gamma_u)_{u\in T}$
\be\label{reparam}
\bbeta=\bA\bgamma,
\ee
where $\bA\in\{0,1\}^{p\times |T|}$  is a tree-induced indicator matrix with entry $A_{jk}=1_{k\in \mbox{\footnotesize Ancestor}(j)\cup \{j\}}$ (or, equivalently, $1_{j\in \mbox{\footnotesize Descendant}(k)\cup \{k\}}$, with $\mbox{Descendant}(k)$ being the set of descendants of node $k$.).

The reparameterization provides a means to couple the tree structure with the coefficients for compositions.
Since the root node value appears in every coefficient in $\bbeta$ by construction, we recommend fixing it to a constant to avoid unnecessary redundancy.
In practice, we set it to be $n^{-1}\1_n^T\by$ to capture the mean of the response {\red where $\1_n$ is a length-$n$ vector of ones}. 
Hereafter, we always assume the intermediate coefficient for the root node is prefixed.
Next, we will introduce regularization on the intermediate coefficients to induce tree-guided equi-sparsity in $\bbeta$.

With the new parameterization, it becomes immediately clear that aggregating features that share the same ancestor node $u$ is equivalent to zeroing out all the intermediate coefficients for nodes in Descendant$(u)$.
For example, in Figure \ref{fig:tree}, if $\gamma_1=\gamma_2=0$, the leaf nodes 1 and 2 sharing the same parent node 8 will be combined and their common coefficient is $\beta_1=\beta_2=\gamma_{10}+\gamma_8+\gamma_{12}$. 
As a result, the desired tree-guided equi-sparsity regularization on $\bbeta$ can be equivalently expressed as structured zero-sparsity regularization on $\bgamma$.
In particular, we propose three sparsity-inducing penalty terms on the intermediate coefficients:
\begin{itemize}
\item[(1)] {\it Node $\ell_1$} (L1):
\be\label{pen1}
\mathcal{P}_{T}(\bgamma)=\sum_{u\in T_{-r}} w_u|\gamma_u|,
\ee
where $T_{-r}$ denotes the set of nodes in $T$ without the root node;
\item[(2)] {\it Child $\ell_2$} (CL2):
\be\label{pen2}
\mathcal{P}_{T}(\bgamma)=\sum_{u\in I(T)} w_u\|(\gamma_v)_{v\in \mbox{\footnotesize Child}(u)}\|,
\ee
where $\mbox{Child}(u)$ denotes the set of children nodes of node $u$;
\item[(3)] {\it Descendant $\ell_2$} (DL2):
\be\label{pen3}
\mathcal{P}_{T}(\bgamma)=\sum_{u\in I(T)} w_u\|(\gamma_v)_{v\in \mbox{\footnotesize Descendant}(u)}\|.
\ee
\end{itemize}
All three penalties induce sparsity in $\bgamma$ and thus potentially result in equi-sparsity in $\bbeta$.
The {\NL} penalty is closely related to the one in \cite{yan2018rare}, except that we do not penalize the original coefficients in $\bbeta$.
This is because the zero-sparsity in $\bbeta$ does not carry any special interpretation in our proposed relative-shift model.
The {\CL} and {\DL} penalties resort to a group-lasso-type regularization which may further encourage the groups of nodes towards the leaves of a tree to take zero values.
In particular, {\CL} does not contain any overlapping groups while {\DL} does.
{Later we show that all three penalty terms can be implemented by the same algorithm and their theoretical properties can be understood through a unified finite-sample prediction error bound.}
The weights in each penalty may be used to adjust for different node heights or heterogeneous group sizes and/or avoid over-penalization if desired. 
By default, we set the weights to be 1 throughout the paper.
Data-adaptive selection of weights is a future research direction.

%
%

Combining with Model \eqref{RS}, we have the following optimization problem
\be\label{opt2}
(\widehat{\bbeta_c},\widehat{\bbeta})=\argmin_{\bbeta_c,\,\bbeta=\bA\bgamma}\quad g(\bbeta_c,\bbeta)+\lambda\pen_T(\bgamma).
\ee
The method utilizes data across all taxonomic ranks and naturally strikes a good balance between data resolution and accuracy.
Moreover, it also achieves taxonomic rank selection by adaptively identifying the most relevant taxa at the most relevant taxonomic ranks.
For example, suppose we start with species-level OTU data and fit a relative-shift model with the taxonomic-tree-guided regularization.
If all the species within a genus are regularized to have the same coefficient, a new taxon is formed at the genus level with its composition being the sum of all the child species compositions, and the genus-level taxon is deemed relevant in the regression analysis rather than its descendant species.
Similarly, if all the species (in different genera) within a family share the same coefficient, the newly formed family-level taxon is selected.

\section{Model Fitting Algorithm}\label{sec:comp}
Both optimization problems in \eqref{opt1} and \eqref{opt2} are convex. In principle, generic convex optimization solvers can be used.
Nonetheless, given the high dimensional nature of the problem, such generic methods are usually computationally prohibitive.
Instead, we resort to a more efficient smoothing proximal gradient\st{ (SPG)} method \citep{chen2012smoothing} to solve the optimization in \eqref{opt1} and \eqref{opt2}.
We remark that the details of the {\SPG} algorithm are well documented in \cite{chen2012smoothing}, so we only outline the general idea of the algorithm below.

The optimization problems in \eqref{opt1} and \eqref{opt2} can be uniformly expressed as
\be\label{opt_uni}
\min_{\widetilde{\bbeta}}\ {1\over 2n}\|\by-\widetilde{\bX}\widetilde{\bbeta}\|^2+\lambda\Omega(\widetilde{\bbeta}),
\ee
where $\widetilde{\bX}=(\bC,\bX)$, $\widetilde{\bbeta}=(\bbeta_c^T,\bbeta^T)^T$, and $\Omega(\widetilde{\bbeta})=\pen_E(\bbeta)$ in \eqref{opt1}, and $\widetilde{\bX}=(\bC,\bX\bA)$, $\widetilde{\bbeta}=(\bbeta_c^T,\bgamma^T)^T$, and $\Omega(\widetilde{\bbeta})=\pen_T(\bgamma)$  in \eqref{opt2}.
In particular, the penalty term $\Omega(\widetilde{\bbeta})$ is a nonsmooth function of $\widetilde{\bbeta}$ and the elements of $\widetilde{\bbeta}$ is nonseparable.
The fundamental idea of {\SPG} is to 1) decouple the nonseparable elements via the dual norm; 2) apply a Nesterov smoothing technique \citep{nesterov2005smooth} to obtain the gradient of $\Omega(\widetilde{\bbeta})$; and 3) apply an optimal gradient method \citep{beck2009fast}.

More specifically, the term $\Omega(\widetilde{\bbeta})$ in \eqref{opt_uni} can be expressed by the dual norm as
\[
\Omega(\widetilde{\bbeta})=\max_{\balpha\in\mathcal{Q}} \balpha^T\bD\widetilde{\bbeta},
\]
where $\mathcal{Q}$ is some convex, closed unit ball and $\bD$ is a constant matrix defined by respective problems (see \cite{chen2012smoothing} for details).
Subsequently, it is approximated by a surrogate function
\be\label{alpha}
f_\mu(\widetilde{\bbeta})=\max_{\balpha\in\mathcal{Q}} {\balpha^T\bD\widetilde{\bbeta}-{\mu\over 2}\|\balpha\|^2},
\ee
which can be shown to be smooth with respect to $\widetilde{\bbeta}$ (as long as $\mu>0$) and bounded by a tight interval around $\Omega(\widetilde{\bbeta})$ \citep{nesterov2005smooth}.
\cite{nesterov2005smooth} further showed that the gradient of $f_\mu(\widetilde{\bbeta})$ is $\bD^T\balpha^*$ with $\balpha^*$ being the optimal solution to \eqref{alpha} and the gradient is Lipschitz continuous.
In particular, in both problems \eqref{opt1} and \eqref{opt2}, $\balpha^*$ has a closed-form expression and the Lipschitz constant is explicit \citep{chen2012smoothing}.

Let $h(\widetilde{\bbeta})={\red (2n)^{-1} \mathst{1\over 2n}}\|\by-\widetilde{\bX}\widetilde{\bbeta}\|^2+\lambda f_\mu(\widetilde{\bbeta})$ be the new objective function.
The gradient of $h(\widetilde{\bbeta})$, i.e., $\triangledown h(\widetilde{\bbeta})$, has an explicit form and is Lipschitz continuous with an explicit Lipschitz constant $L$.
To minimize $h(\widetilde{\bbeta})$, one may resort to the classical gradient algorithm by iteratively updating the estimate of $\widetilde{\bbeta}$:
\[
{\widetilde{\bbeta}}^{(t+1)}={\widetilde{\bbeta}}^{(t)}-{1\over L} \triangledown h({\widetilde{\bbeta}}^{(t)}),
\]
until convergence.
However, the convergence may be slow \citep{nesterov1983method}.
Instead, {\SPG} applies the fast iterative shrinkage-thresholding algorithm\st{ (FISTA)} \citep{beck2009fast} which is an optimal gradient method in terms of convergence rate.
{\red The fast iterative shrinkage-thresholding algorithm\st{FISTA}} updates the estimate ${\widetilde{\bbeta}}^{(t+1)}$ with not just the previous estimate ${\widetilde{\bbeta}}^{(t)}$, but rather a very specific  combination of the previous two estimates ${\widetilde{\bbeta}}^{(t)}$ and ${\widetilde{\bbeta}}^{(t-1)}$.
As a result, the convergence has been proved to be much faster than the standard gradient method \citep{chen2012smoothing,beck2009fast}.

The tuning parameter $\lambda$ in \eqref{opt_uni} balances the quadratic loss function and the penalty term. In practice, it typically has to be determined from data. A standard approach is to use cross validation to adaptively select the optimal tuning parameter. Since the {\SPG} algorithm for model fitting is very efficient, the cross validation scheme is computationally feasible. We provide more details in the numerical studies in Section \ref{sec:sim}.

\section{Theory}\label{sec:theory}
Let $T$ represent a $p$-leafed taxonomic tree with root node $r$.
Both $L(T)$ and $I(T)$ have been defined previously as the sets of leaf nodes and internal nodes, respectively.
Let $T_u$ be a subtree of $T$ rooted at the node $u$ for $u \in T$. 


To focus on the main idea, we consider the relative-shift model without additional covariates,
\be\label{RS1}
\by = \bX\bbeta^* + \bvare,
\ee
where $\bbeta^*\in\real^p$ is the true coefficient vector.
With the tree-based reparameterization \eqref{reparam}, we have
\[
\bbeta^*=\bA\bgamma^*,
\]
where  $\bgamma^*=(\gamma_u^*)_{u\in T}$ is the vector of intermediate coefficients and $\bA$  
is a tree-induced indicator matrix.
Without loss of generality, we assume the response $\by$ is centered and the root node takes value 0 (i.e., $\gamma_r=0$).
Correspondingly, $\bgamma^*=(\gamma_u^*)_{u\in T\backslash \{r\}}\in\mR^{|T|-1}$ and $\bA\in\{0,1\}^{p\times (|T|-1)}$.

We study finite-sample properties of the regularized estimator
\begin{eqnarray} \label{eq:tree-pen}
\widehat{\bbeta} = \argmin_{\bbeta=\bA \bgamma}\left\{ \frac{1}{2n} \|\by - \bX \bbeta\|^2 + \lambda \mathcal{P}_{T}(\bgamma)\right\},
\end{eqnarray}
where $\mathcal{P}_{T}(\bgamma)$ is any one of the three penalties (i.e., {\NL}, {\CL}, and {\DL}) introduced in Section \ref{subsec:tax}.
We formally state the assumptions on the design matrix and the random error vector and present our main results in Theorem \ref{eq:th1}. The detailed proof is in {the Section~\ref{app:proofs} of Supplementary Materials}.


\begin{assumption}\label{assumption1}
The entries of  $\bvare$ are independently and identically drawn from standard Gaussian distribution with mean zero and variance $\sigma^2$.
\end{assumption}

\begin{assumption}\label{assumption2}
The design matrix $\bX\in \mathbb{R}^{n\times p}$ is compositional, i.e., each entry of $\bX$ is in the interval $[0,1]$ and each of its row sums up to 1.
\end{assumption}

\begin{theorem}\label{eq:th1}
Suppose Assumptions \ref{assumption1}--\ref{assumption2} hold. Consider the regularized estimator $\widehat{\bbeta}$ of $\bbeta$ from solving \eqref{eq:tree-pen} with any penalty forms in \eqref{pen1}--\eqref{pen3}. Denote $|I(T)|$ as the number of internal nodes of the tree. Choose $\lambda \ge 2 \sqrt{2}\sigma\sqrt{\log(|I(T)|) /(\delta n)}$. Then with probability at least $1-\delta$, it holds that
\begin{equation}\label{eq:bound}
  \frac{1}{n}\|\bX \widehat{\bbeta} - \bX \bbeta^*\|^2 \preceq 
  \lambda \{\min_{\bgamma; \bA\bgamma = \bbeta^*} \mathcal{P}_T(\bgamma)\},
\end{equation}
where $\preceq$  means the inequality holds up to a multiplicative constant.
\end{theorem}



In the above results, the order of $\lambda$ is $O(\sqrt{\log(|I(T)|)/n})$, depending on the tree structure through the total number of internal nodes $|I(T)|$ that represents the dimension of the model. The term $\{\min_{\bgamma; \bA\bgamma = \bbeta^*} \mathcal{P}_T(\bgamma)\}$ captures the complexity of the true model by measuring the minimal penalty function evaluated at the truth.

With the above unified prediction error bound, we now perform further analysis on mode size and complexity to obtain specific error rates. Following \citet{yan2018rare}, we first introduce the concepts of aggregating set and coarsest aggregating set, which correspond to the equi-sparsity pattern of the coefficients in the proposed relative shift model.

\begin{definition}\label{def1}
  We say that $B \subseteq T$ is an aggregating set with respect to
  $T$ if $\{L(T_u): u \in B\}$ forms a partition of $L(T)$.
\end{definition}

\begin{definition}\label{def2}
  For any $\bbeta^{*} \in \mathbb{R}^p$, there exists a unique
  coarsest aggregating set $B^* := B(\bbeta^{*}, T) \subseteq T$
  (``the aggregating set'') with respect to the tree $T$ such
  that (a) $\beta^{*}_j = \beta^{*}_k$ for $j, k \in L(T_u)$
  $\forall u \in B^*$, (b) $|\beta^{*}_j - \beta^{*}_k| > 0$ for
  $j \in L(T_u)$ and $k \in L(T_v)$ for siblings $u, v \in B^*$.
\end{definition}


As an example, consider the tree structure in Figure~\ref{fig:tree}.
Assume $\beta^*_1=\beta^*_2$ and $\beta_5^*=\beta^*_6=\beta_7^*$, and all other coefficients are distinct.
The node set $\{3,4,8,11\}$ forms an aggregating set
since the leaves of the four subtrees rooted by the four nodes form a partition of the leaf nodes $\{1, \ldots, 7\}$. Moreover, this node set is also the coarsest aggregating set corresponding to the equi-sparsity pattern of $\bbeta^*$.

It is then clear that the size of the coarsest aggregating set,
$|B^*|$, is a natural complexity measure of the equi-sparsity pattern of $\bbeta^*$ guided by the tree. Therefore, it is desired to bound $\{\min_{\bgamma; \bA\bgamma = \bbeta^*} \mathcal{P}_T(\bgamma)\}$
in Theorem \ref{eq:th1} in terms of $|B^*|$ and the magnitude of $\bbeta^*$.
In particular, for {\NL} in \eqref{pen1} and {\CL} in \eqref{pen2}, we have the following lemma (see  Section~\ref{app:proofs} of Supplementary Materials for a detailed proof).
\begin{assumption}\label{assumption3}
$T$ is a $p$-leafed full tree such that each node is either a leaf or possesses at least two child nodes.
\end{assumption}

\begin{assumption}\label{assumption4}
The true coefficient $\bbeta^*$ is bounded, i.e., $\|\bbeta\|_{\infty} \leq M$, where $M>0$ is a constant.
\end{assumption}

\begin{lemma}\label{lemma:penalty}
  Suppose Assumptions \ref{assumption3}--\ref{assumption4} hold.
  For {\NL} in \eqref{pen1} and {\CL}  in \eqref{pen2}, respectively, it holds that
  $$
\min_{\bgamma; \bA\bgamma = \bbeta^*} \mathcal{P}_T(\bgamma) \leq M|B^*|.
  $$
\end{lemma}

Together with Theorem \ref{eq:th1}, we have the following results.

\begin{corollary}\label{corollary:bound}
  Suppose Assumptions \ref{assumption1}--\ref{assumption4} hold. Consider the regularized estimator $\widehat{\bbeta}$ of $\bbeta$ from solving \eqref{eq:tree-pen} with either penalty form in \eqref{pen1} and \eqref{pen2}. Let $B^*$ be the coarsest aggregating set according to $(T,\bbeta^*)$. Choose $\lambda \ge 2 \sqrt{2}\sigma\sqrt{\log(p) /(\delta n)}$. Then with probability at least $1-\delta$, it holds that
\begin{equation}\label{eq:bound2}
  \frac{1}{n}\|\bX \widehat{\bbeta} - \bX \bbeta^*\|^2 \preceq 
  \sigma\sqrt{\log(p) /(\delta n)}|B^*|.
\end{equation}
\end{corollary}

{We remark that the bound takes a familiar form as those for many well-studied high-dimensional models. Due to Assumption \ref{assumption3}, the measure of the model dimension, i.e., the number of internal nodes $|I(T)|$, is of order $p$. The $\log(p)$ term then represents the price we have to pay due to high dimensionality. Both {\NL} and {\CL} can predict well as long as $\log(p)/n = o(1)$ and its performance is tied to $|B^*|$, representing the complexity of the equi-sparsity pattern on the tree.}


\section{Simulation}\label{sec:sim}
\subsection{\red Preliminaries}
We compare the proposed relative-shift regression framework with the transformation-based regression models using comprehensive simulations.
Specifically, we consider the relative-shift model with the equi-sparsity regularization (i.e., ``RS-ES") and the three tree-guided regularization methods, {\red {\it Node $\ell_1$}, {\it Child $\ell_2$} and {\it Descendant $\ell_2$} (denoted as ``RS-L1", ``RS-CL2", and ``RS-DL2" respectively), when applicable}.
For competing methods, we consider the log-contrast model with lasso penalty (``LC-Lasso") \citep{lin2014variable} and the kernel penalized regression (KPR) model \citep{randolph2018kernel} with ridge kernel (``KPR-Ridge") and taxonomic kernel (``KPR-Tree").
Since the proposed relative-shift paradigm is fundamentally different from the log-contrast models, there is no point of directly comparing parameter estimation accuracy.
Instead, we focus on the comparison of prediction accuracy and computing times under various generative models.
Each simulation study is repeated 100 times.
All tuning parameters are selected using cross validation.

\subsection{Study I: Equi-Sparsity Setting}

In this study, we first simulate relative abundance data $\bX$ for $p=100$ taxa and $n=500$ samples (i.e., $100$ for training and $400$ for testing).
In particular, the compositions  are generated from a logistic Gaussian distribution where we first simulate a Gaussian data matrix $\bZ=(z_{ij})$ of size $n\times (p-1)$ and then obtain the compositional vector
$\bx_{i}=\left(e^{z_{i1}} / \{1+\sum_{j=1}^{p-1}e^{z_{ij}}\},\cdots,\right.$ $\left.e^{z_{i,p-1}}/ \{1+\sum_{j=1}^{p-1}e^{z_{ij}}\},1 / \{1+\sum_{j=1}^{p-1}e^{z_{ij}}\}\right)^T$
for the $i$th subject.
The resulting relative abundance matrix $\bX$ does not have zero values.
To further mimic the reality, we truncate the data by 0.005, which leads to about 40\% zero entries.
Then we renormalize the data to be compositions and denote the zero-inflated matrix as $\bX_0$.
In what follows, we assume the true model is generated from $\bX$ but we only use $\bX_0$ in model fitting and testing. 

Then we simulate response from the relative-shift Model \eqref{RS}, where the coefficients for compositions are equi-sparse with the first twenty entries being -1, the next ten entries being 2, and remaining being 0.
Namely, from the prediction perspective, Taxa 1-20, Taxa 21-30, and Taxa 31-100 can be aggregated, respectively, without losing any information.
We further simulate random errors with variance such that the signal-to-noise ratio (SNR) is 1.

\begin{figure}[hbpt]
\centering
       \includegraphics[width=2.5in]{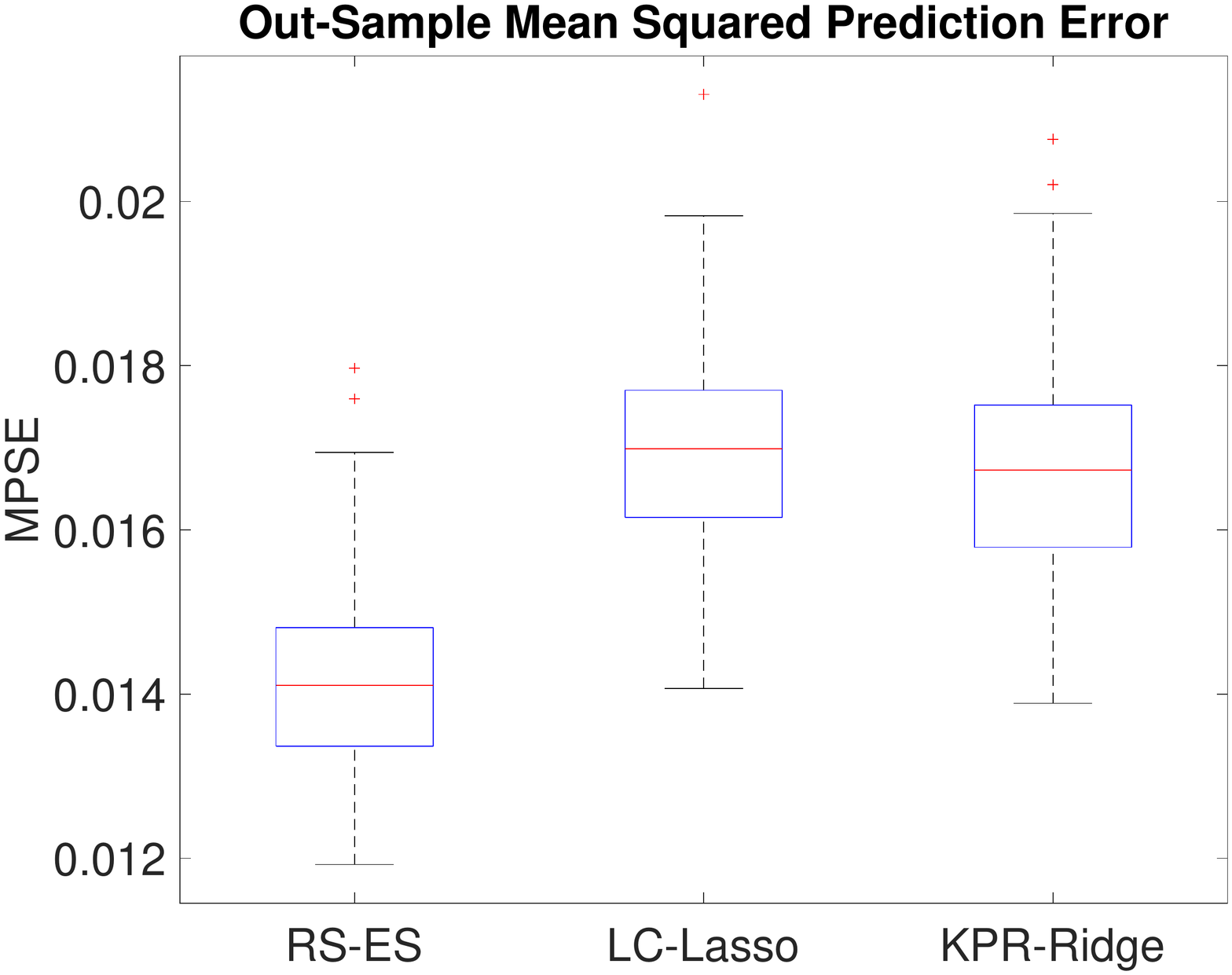}
       \includegraphics[width=2.5in]{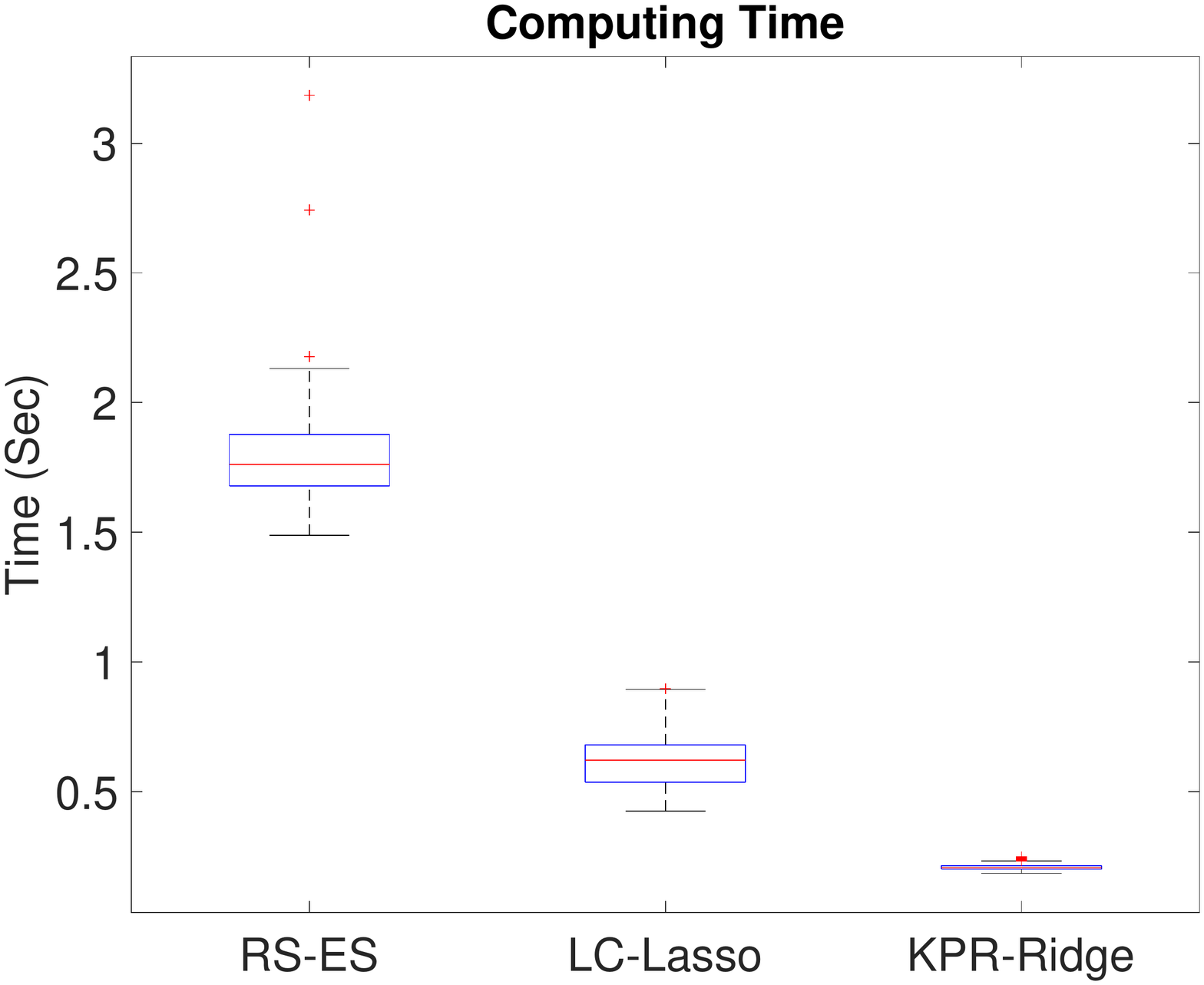}
\caption{MPSE and computing time comparison in simulation Study I.}
\label{fig:sim1}
\end{figure}

Since there is no extrinsic taxonomic tree information in this setting, we just compare RS-ES, LC-Lasso, and KPR-Ridge.
In particular, since both LC-Lasso and KPR-Ridge rely on logarithmic transformations which cannot handle zero, we follow the convention and substitute zero values with a preset small value (i.e., $0.001$) and renormalize data.
The out-sample mean squared prediction error (MSPE)
\bes
MSPE={1\over 400} {\sum_{i=1}^{400} (y_i-\widehat{y}_i)^2}
\ees
and the computing time (including cross validation for tuning parameter selection) are assessed for each method.
The comparison can be found in Figure \ref{fig:sim1}.
Apparently, RS-ES significantly outperforms the two transformation-based methods in prediction accuracy.
All three methods are computationally efficient as the model fitting times are within a couple of seconds on a standard desktop computer (16Gb RAM, Intel Core i7 CPU 2.20 GHz).
We remark that the MPSE of the competing methods may change as we vary the preset value of the zero surrogate.
This indicates that the transformation-based methods may be unstable in handling data with excessive zeros.

We also compare different methods when data are generated from a log-contrast model.
The proposed method is comparable to the log-contrast counterparts in this misspecified setting, especially when the SNR is low.
Detailed results are contained in {Section~\ref{app:add_sim} of Supplementary Materials}.

\subsection{Study II: Tree-Guided Equi-Sparsity Setting}
\begin{figure}[hbpt]
\centering
\subfloat{\includegraphics[width=.5\textwidth]{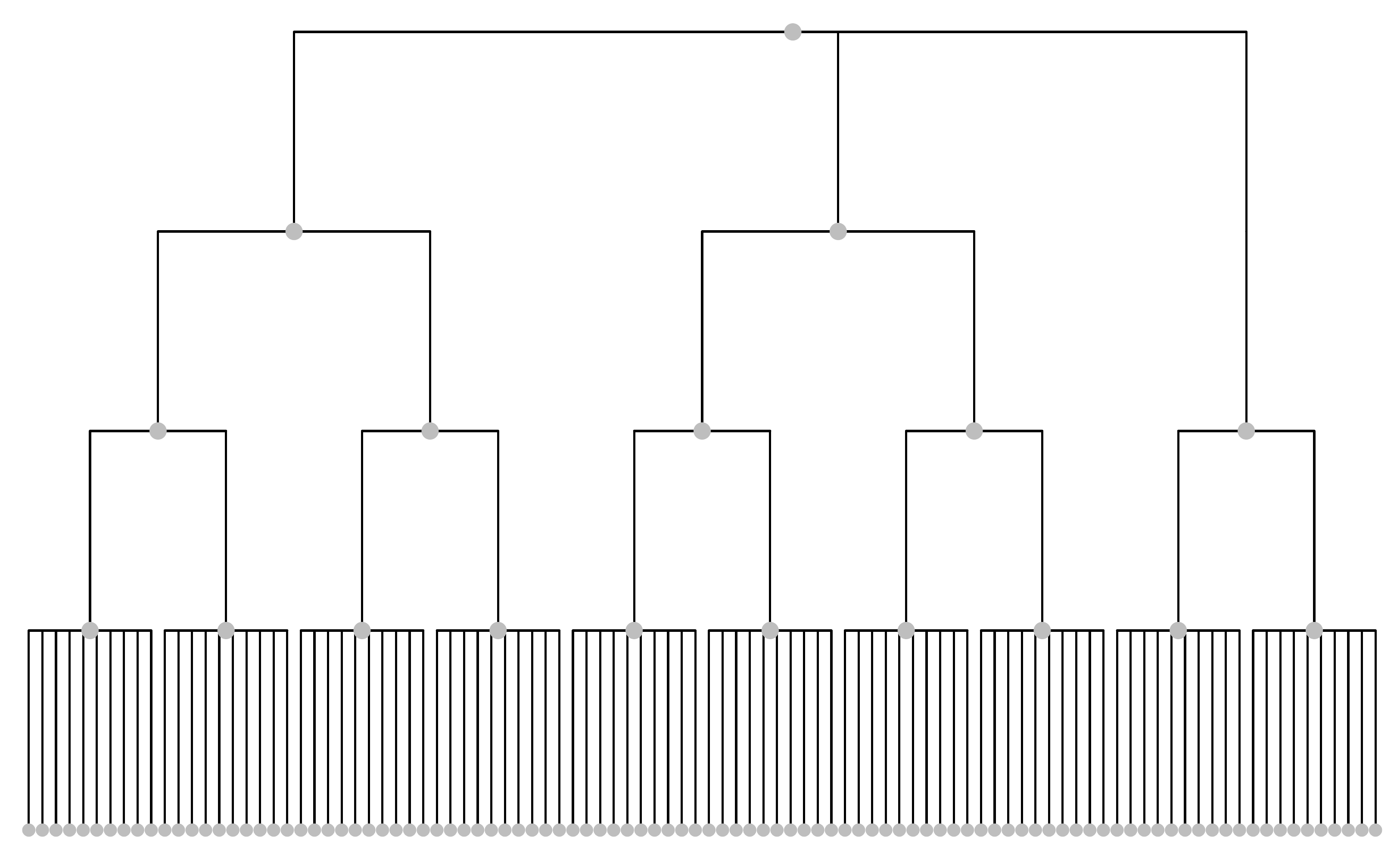}}
\subfloat{\includegraphics[width=.5\textwidth]{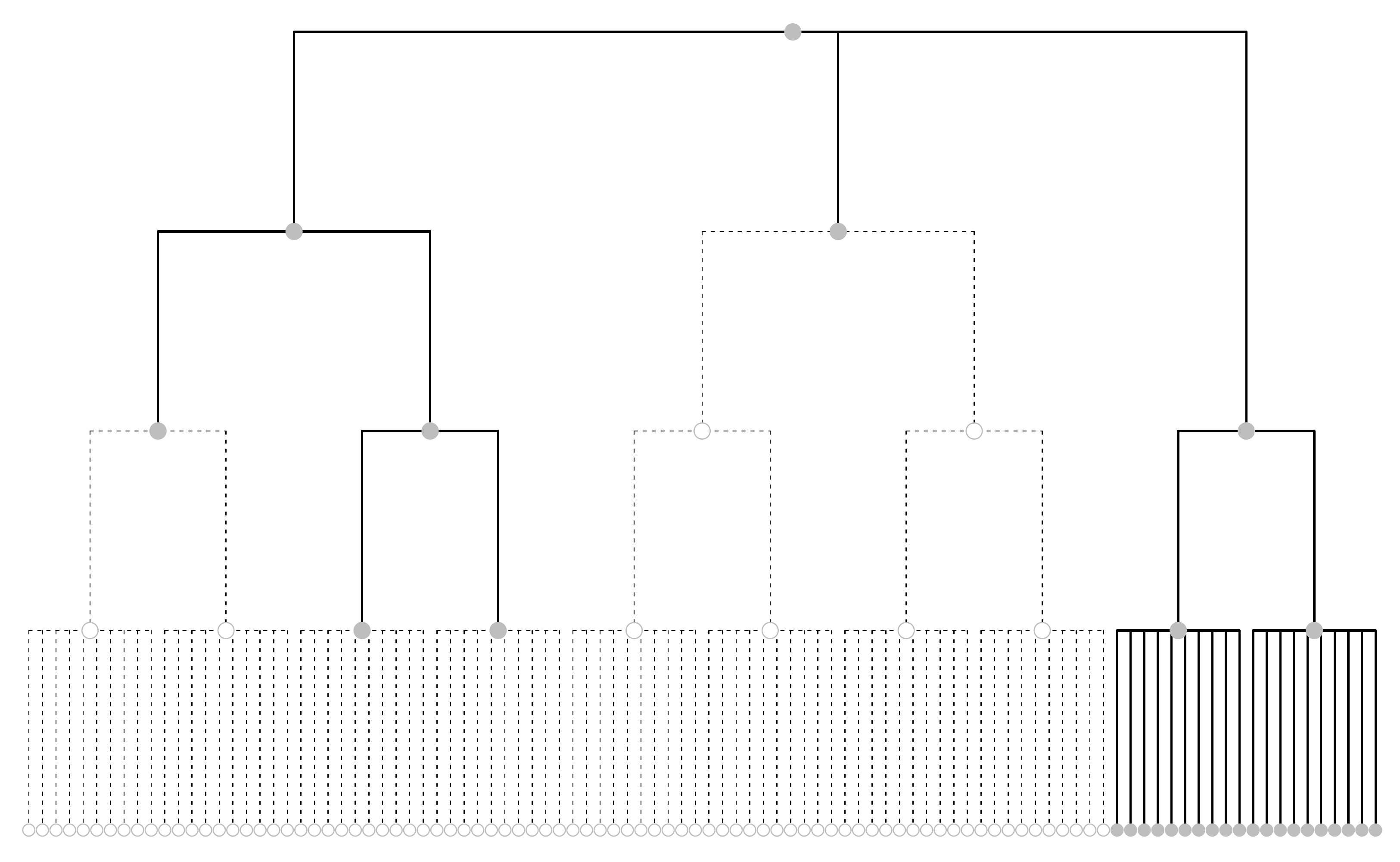}}
\caption{Left: The taxonomic tree structure among variables in Study II. The leaf node indices are in ascending order from left to right. Right: The equi-sparsity structure of the regression coefficients. Features with the same coefficient are aggregated to the common ancestor (i.e., the closest solid node).
}
\label{fig:simtree}
\end{figure}

In Study II, we include  extrinsic information of a taxonomic tree.
The compositional data are generated in the same way as in Study I.
In particular, we assume there is a tree structure among the $p=100$ variables as shown in the left panel of Figure \ref{fig:simtree}, where every 10 consecutive leaf nodes share a common parent node and so on.
Guided by the tree structure,  the true coefficient vector for the generative relative-shift model is set to be
\[
\bbeta=\left(\1_{20}^T,\ -2\cdot\1_{10}^T,\ 0.5\cdot\1_{10}^T,\ 2\cdot\1_{40}^T,\ \bxi_{20}^T\right)^T,
\]
where $\1_q$ is a length-$q$ vector of ones and $\bxi_{20}$ is a length-20 vector filled standard Gaussian random numbers.
Let us define the level of a node to be the number of connections between the node and the root plus 1.
The coefficients indicate that the first 20 variables on level 5 are aggregated to the internal node on level 3; the next 10 variables on level 5 are aggregated to the internal node on level 4, and so on.
The last 20 variables have distinct values, meaning that they cannot be further aggregated.
The feature aggregation is shown in the right panel of Figure \ref{fig:simtree}.

\begin{figure}[hbpt]
\centering
       \includegraphics[width=2.5in]{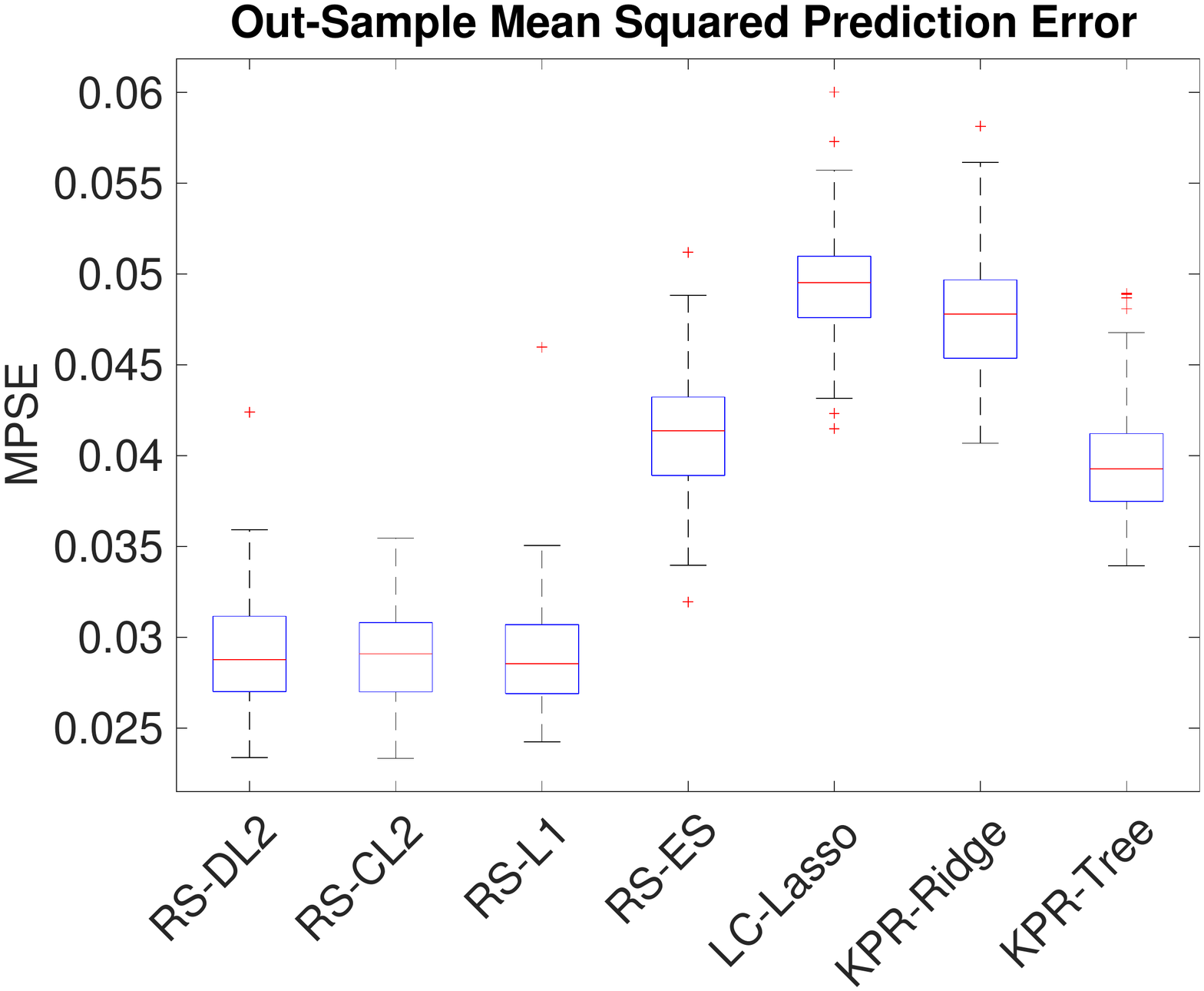}
       \includegraphics[width=2.5in]{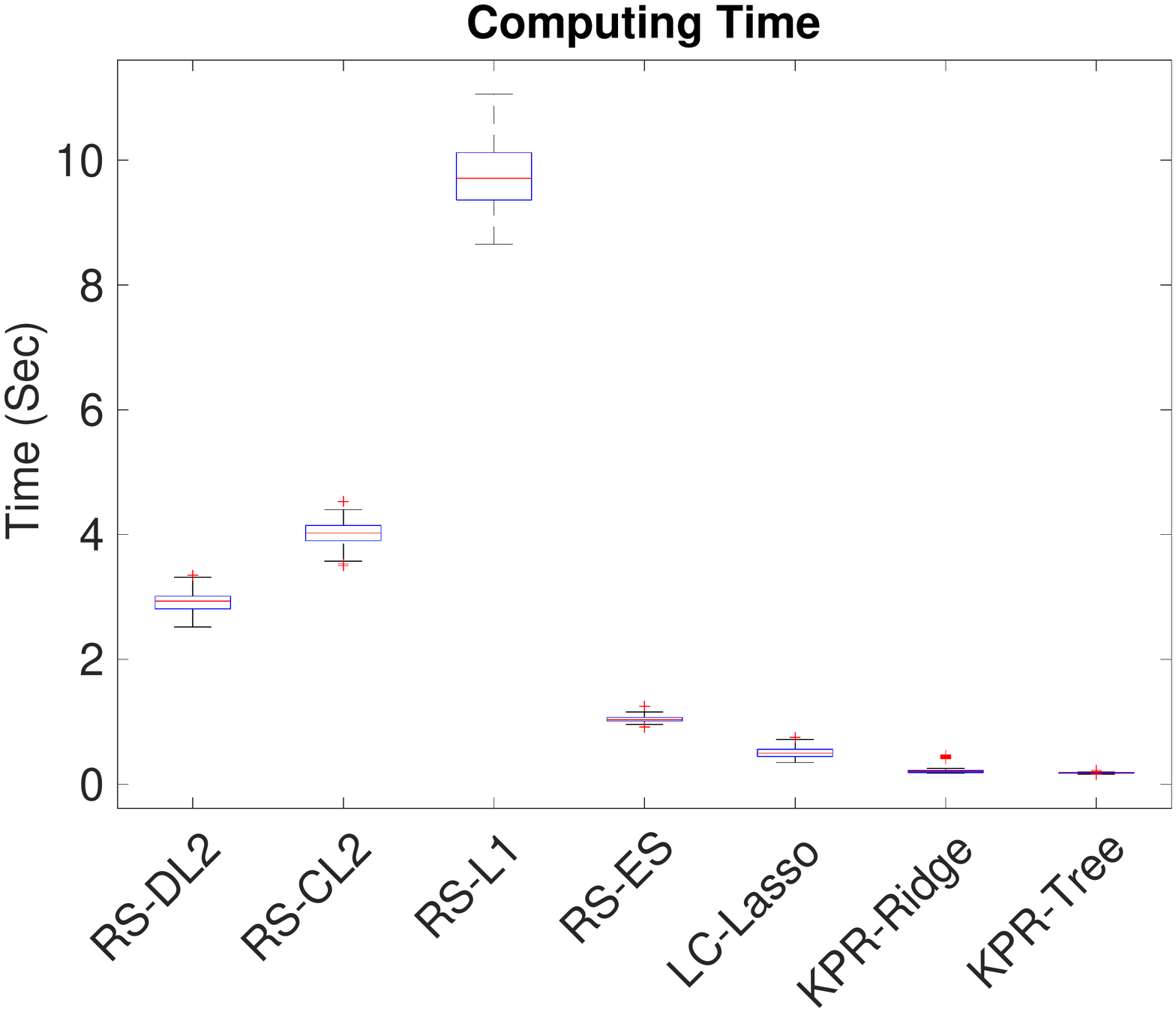}
\caption{MPSE and computing time comparison in simulation Study III.}
\label{fig:sim3}
\end{figure}

We apply the three tree-guided regularization methods RS-L1, RS-CL2, and RS-DL2, as well as RS-ES, LC-Lasso, KPR-Ridge, and KPR-Tree to the data.
We remark that other than the three proposed methods, only KPR-Tree can take advantage of the tree structure.
In particular, KPR-Tree converts the tree structure to a patristic distance kernel.
In addition, we substitute zero values by a small preset value for LC-Lasso, KPR-Ridge, and KPR-Tree just as in Study I.
The comparison is shown in Figure \ref{fig:sim3}.
In the left panel, the three tree-guided relative-shift methods perform similarly and are significantly better than the other methods in terms of the prediction accuracy.
The next best method is KPR-Tree, which also takes into account the guidance from the extrinsic tree structure.
RS-ES is slightly worse than KPR-Tree, but significantly better than LC-Lasso and KPR-Ridge.
On the other hand, the superior prediction performance of the tree-guided relative-shift methods does come at a price, that is, a slightly higher computational cost (especially for RS-L1).
However, considering the scale of the problem and the cross validation scheme for tuning parameter selection, the computing time is quite acceptable.

We also conduct additional simulations with various zero proportions and SNRs. The results are largely consistent with what has been reported here.
One thing we find intriguing is that the three proposed tree-guided regularization methods almost always have similar prediction performance.
We further investigate the three methods in a separate study in {Section~\ref{app:add_sim} of Supplementary Materials}.
In general, the three methods only differ slightly in the estimation of the intermediate coefficients.
They almost always have similar estimation (of the original coefficients) and prediction performance, and thus may be used exchangeably in practice.
A more comprehensive comparison between different regularization methods is left for future research.


\section{Application to Preterm Infant Gut Microbiome Study}\label{sec:real}
We apply the proposed relative-shift model with taxonomic-tree-guided regularization to a preterm infant gut microbiome study.
The study aims to investigate how gut microbiome is related to the neurodevelopment of preterm infants.
Data were collected at a Neonatal Intensive Care Unit (NICU) in the northeast US. Fecal samples of preterm infants were collected daily when available during the infant’s first month of postnatal age.
Bacterial DNA was isolated and extracted from each sample \citep{Bomar2011,cong2017influence}; V4 regions of the 16S rRNA gene were sequenced using the Illumina platform. Gender, birth weight, delivery type, and complications were recorded at birth, and medical procedures and feeding types were recorded throughout the infant’s stay. Infant neurobehavioral outcomes were measured when the infant reached 36-38 weeks of postmenstrual age, using the NICU Network Neurobehavioral Scale (NNNS) \citep{cong2017influence}.

\begin{figure}[hbpt]
\centering
       \includegraphics[width=5in]{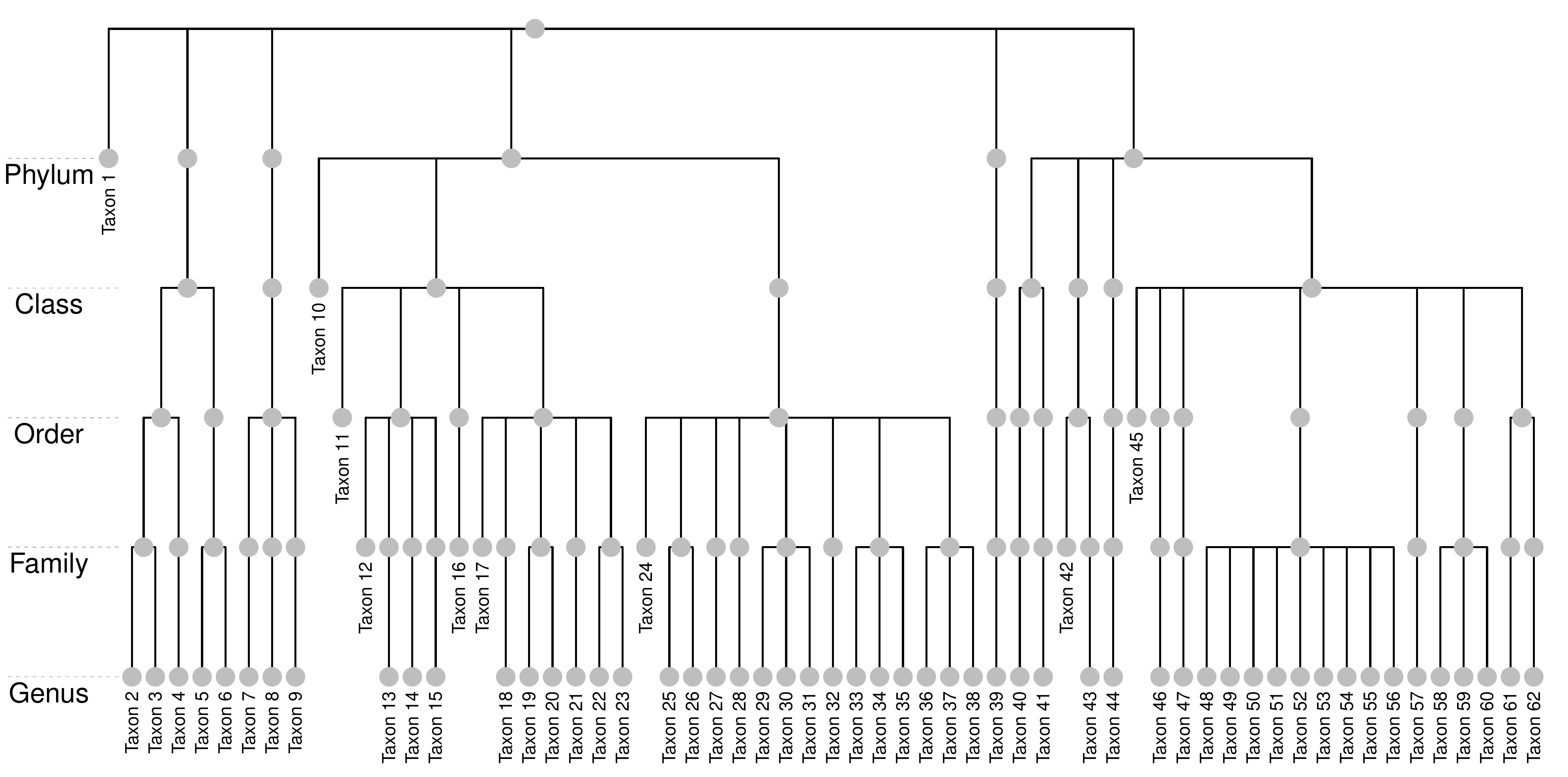}
\caption{Taxonomic tree of the 62 taxa in the NICU data.}
\label{fig:realtree}
\end{figure}

To obtain the OTU table for analysis, we cluster and analyze the raw data using {\red Quantitative Insights Into Microbial Ecology}\st{QIIME} \citep{QIIME2010}.
The data are classified up to the genus level.
After proper processing, we obtain $p=62$ taxa, most at the genus level, on $n=34$ individuals.
The longitudinal data are averaged across the postnatal period for each infant, resulting in a single $34\times 62$ compositional matrix with 39.2\% zero entries.
Moreover, the taxonomic tree of the 62 taxa is also available (see Figure \ref{fig:realtree}).
Each taxon in the OTU table corresponds to a leaf node.
Most taxa are at the genus level while some are at higher levels.
The primary outcome is the continuous NNNS score, and we include 6 additional covariates (i.e., gender, delivery type, premature rupture of membranes, score for Neonatal Acute Physiology–Perinatal Extension-II (SNAPPE-II), birth weight, and percentage of feeding with mother’s breast milk) in our analysis.

\begin{figure}[hbpt]
\centering
       \includegraphics[width=5in]{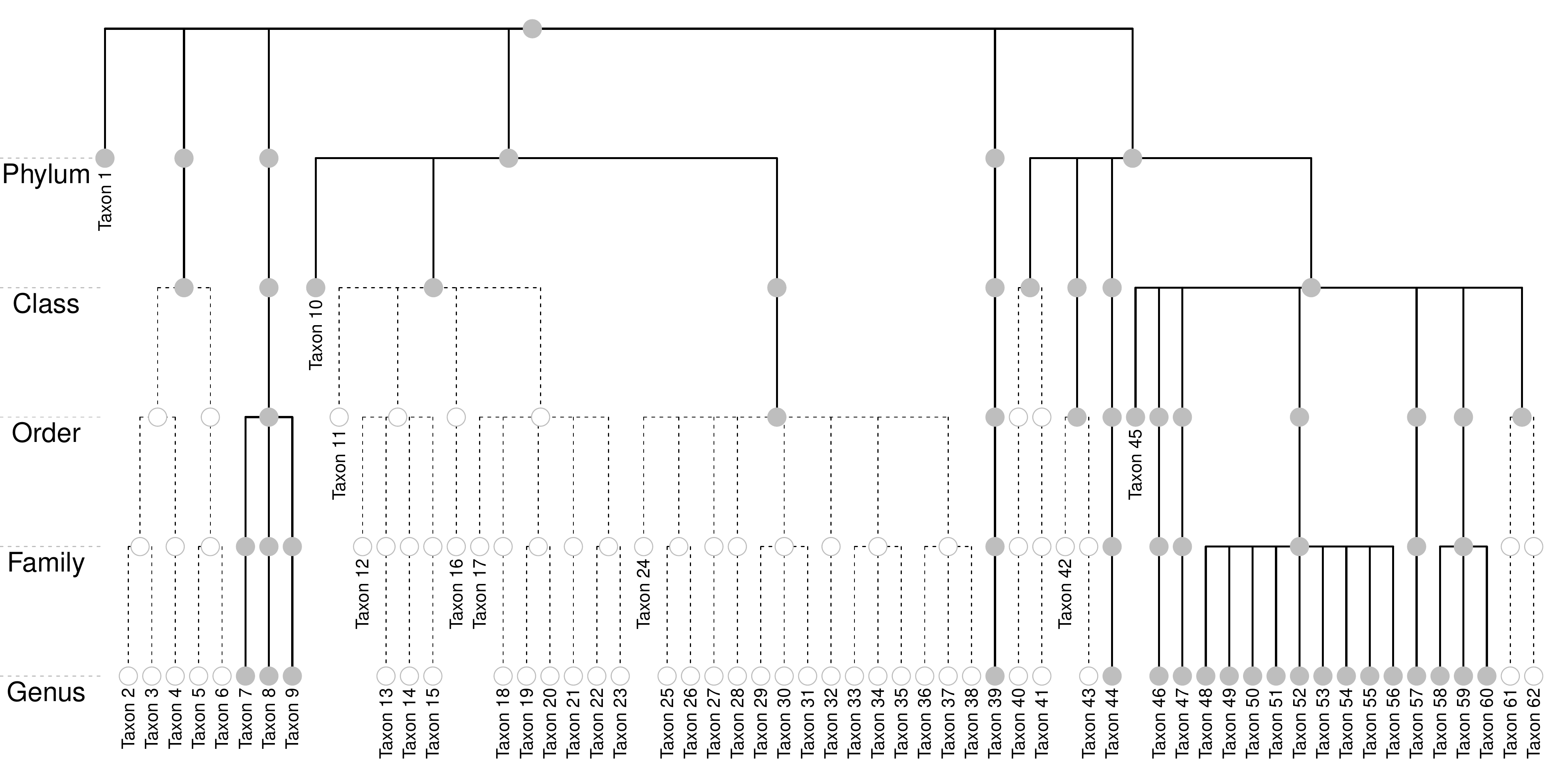}
\caption{Estimated RS-DL2 coefficients for compositions in the NICU data. Taxa with the same title color have the same coefficient and are aggregated to common ancestor (i.e., the closest solid node). }
\label{fig:coef}
\end{figure}

We apply RS-DL2 to the NICU data with covariate adjustment (both RS-CL2 and RS-L1 lead to very similar results and thus are omitted).
The tuning parameter is chosen by 5-fold cross validation.
The estimated coefficients for compositions are approximately equi-sparse but not exact.
This is a common issue with the group-lasso-type penalty \citep{chen2012smoothing}.
To facilitate interpretation, we set a small threshold (i.e., $10^{-4}$) and truncate the groups of intermediate coefficients whose Frobenius norms are below the threshold.
As a result, we obtain highly interpretable equi-sparse coefficients for the 62 taxa; an illustration of the estimated coefficients for compositions on the taxonomic tree is provided in Figure \ref{fig:coef}. 
Blank nodes have zero intermediate coefficients.
Taxa with the same coefficient are aggregated to the common ancestor (i.e., the closest solid node).
For instance, Taxa 2-6 have the same coefficient, indicating that the total composition of these five taxa (at the class level) { is what matters to the prediction of the outcome.}

The findings are largely consistent with our previous result at the order level \citep{sun2018log}, where Lactobacillales (Taxa 17-23), Clostridiales (Taxa 24-38), Enterobacteriales (Taxa 48-56), and other unclassified bacteria (Taxon 1) are identified to be significantly associated with the stress score in infants.
The proposed method provides more insightful results with better biological interpretation.
In particular, Figure \ref{fig:coef} shows that composition of the order Clostridiales (Taxa 24-38) as a whole matters \citep{sordillo2019association} rather than its child taxonomies at the family or genus level. In contrast, the order Enterobacteriales (Taxa 48-56) is important because the compositions of all genera in this order are relevant.
This is very intriguing because higher abundance of gut Enterobacteria is a characteristic pattern of dysbiosis \citep{zeng2017mechanisms,degruttola2016current}, and our results thus warrant further analysis at the genus level on how such dysbiosis contributes to neurodevelopmental deficits. In addition, our method also identifies  other relevant taxa at different taxonomic levels, which serve as a basis for further clinical research.
For example, the class Actinobacteria (Taxa 2-6) is selected, which is involved in lipid metabolism \citep{painold2019step} and has been shown to be associated with mood disorders \citep{huang2019current}.

We further apply RS-ES and the competing methods (LC-Lasso, KPR-Ridge, and KPR-Tree) to the NICU data.
We conduct a leave-one-out cross validation (LOOCV) to compare the prediction accuracy of different methods. The tuning parameters for different methods are selected by CV on the training samples in each run.
The prediction squared errors (PSE) are summarized in Table \ref{tab:real}.
We remark that due to the small sample size ($n=34$), all methods have similar performance except for the LC-Lasso method which is apparently inferior to others.
This is mainly because LC-Lasso  cannot properly adjust for covariates, and all coefficients (for covariates and for log-transformed compositions) are equally penalized.
Although the PSE of our methods is slightly higher than that of KPR, the superior interpretability of the estimated coefficients  further warrants the use of the new regression framework in this application.

\begin{table}[htbp]
  \centering
  \caption{The median (median absolute deviation) of PSE (multiplied by $10^3$) of different methods based on LOOCV of the NICU data  }
  \begin{tabular}{|c||c|c|c|c|c|}
    \hline
    Method & RS-DL2 & RS-ES & LC-Lasso & KPR-Ridge & KPR-Tree \\
    \hline
    PSE$*10^3$ & 4.47 (3.49) & 4.44 (3.85) & 6.81 (5.52) & 4.58 (3.95) & 4.18 (3.81) \\
    \hline
  \end{tabular}
\label{tab:real}
\end{table}

\section{Discussion}\label{sec:dis}
In this paper, we develop a novel relative-shift regression paradigm for microbiome compositional data.
The new framework regresses the response on compositions directly without transformation.
As a result, it adequately handles the unique features of microbiome data such as  compositionality and zero inflation.
Moreover, the regression coefficients carry straightforward biological interpretation, that is, the contrast of the regression coefficients can be interpreted as the effect of certain shifts of abundances of a group of taxa while holding their sum fixed.
The relative-shift framework provides a flexible basis for supervised dimension reduction.
We develop different regularization methods, i.e., the equi-sparsity regularization and the taxonomic-tree-guided regularization, for feature aggregation.
In particular, the tree-guided regularization takes advantage of the extrinsic taxonomic structure among taxa and adaptively identifies relevant taxa at different taxonomic levels.
An efficient {\SPG} algorithm is devised to fit models with different regularization terms.
Numerical studies demonstrate that the proposed methods provide an effective and highly interpretable alternative for microbiome regression, especially in low-signal scenarios.

There are a few directions for future research.
First, same as the log-contrast models, the relative-shift model is a linear regression model. In practice, the effect of microbial concentration shifts on the response may be nonlinear. It is of particular interest to generalize the current framework to accommodate such nonlinear relationships. 
{Second, in the theoretical analysis, thus far we are not able to show that the prediction error of the {\DL} estimator can achieve the same order as those of the other two estimators. We will thoroughly investigate this issue in view of their similar numerical performance.} Third, microbiome data are typically measured with errors. In particular, data at lower taxonomic ranks are more granular but less accurate.  Although the proposed taxonomy-guided regularization method can strike a balance between data accuracy and resolution by using data across the tree, tailored error-in-variable methods are warranted to better account for measurement errors. This is especially relevant when the taxonomy is not readily available. Last but not the least, longitudinal microbiome studies are burgeoning as exemplified by the NICU study. There have been some recent developments on longitudinal regression methods for microbiome data \citep{sun2018log}.
The generalization of the relative-shift framework to the longitudinal setting calls for more investigation.

\section*{Acknowledgement}
The authors thank Dr.\ Xiaomei Cong for providing data from the NICU study (supported by U.S. National Institutes of Health Grant K23NR014674) and offering helpful discussions.
Gen Li's research was partially supported by the National Institutes of Health grant R03DE027773.
Kun Chen's research was partially supported by the National Science Foundation grant IIS-1718798.


\newpage
\appendix

\renewcommand{\thefigure}{S.\arabic{figure}}
\renewcommand{\thetable}{S.\arabic{table}}
\renewcommand\theequation{S.\arabic{equation}}
\setcounter{equation}{0}
\newtheorem{proposition}{Proposition}

\section{Additional Simulation Studies}\label{app:add_sim}
\subsection{Log-Contrast Setting}
We replicate the simulation setting in \cite{lin2014variable} and simulate data from a log-contrast model with sparse coefficients.
Same as the Study I in the main paper, we set $p=100$ and $n=500$ (with one fifth being training samples).
More specifically, the compositions are generated from a logistic normal distribution.
Slightly different from Study I, the base Gaussian matrix $\bZ$ has an exponential correlation structure and nonzero mean.
In addition, the true compositional data matrix $\bX$ is assumed known with no zeros.
The true coefficient vector is $\bbeta=(1,-0.8,0.6,0,0,-1.5,-0.5,1.2,0,\cdots,0)^T$.
Namely, there are only 6 out of 100 nonzero coefficients.

\begin{figure}[hbpt]
\centering
       \includegraphics[width=4in,height=3in]{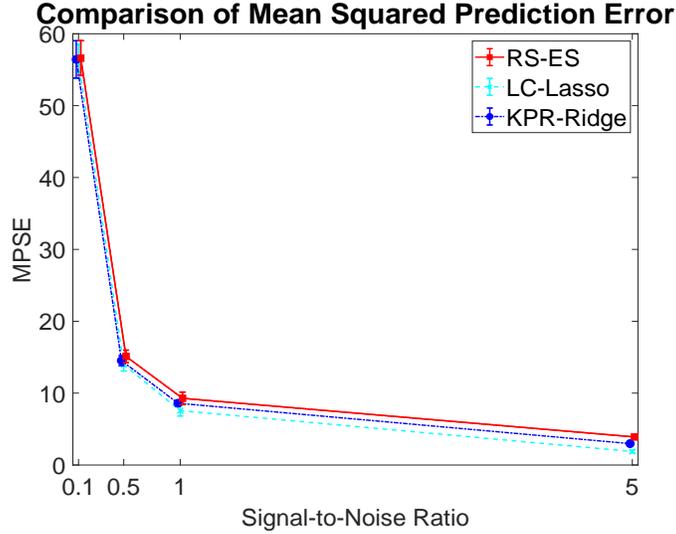}
\vskip-.2in
\caption{Comparison of MPSE over a range of SNRs in simulation Study II. The median and median absolute deviation (error bar) are shown for each method in each scenario. }
\label{fig:sim2}
\end{figure}

In this setting, we set the SNR to be 0.1, 0.5, 1, and 5, and compare RS-ES with LC-Lasso and KPR-Ridge.
The comparison of MPSE is shown in Figure \ref{fig:sim2}.
In particular, LC-Lasso is the best among the three when the SNR is large.
This is probably because LC-Lasso is the true generative model and there is no zero value in $\bX$.
As the SNR decreases, the performance of all methods gets worse as expected, and the difference between different methods becomes negligible.
In other words, the proposed relative-shift method is comparable to the transformation-based methods in low-signal cases even when the true generative model is based on transformed data. 
In reality, data are usually noisy and inflated with zeros.
In those scenarios, the proposed method may have an edge over the log-contrast models.
The computing times of different methods are all within a few seconds and thus the comparison is omitted.

\subsection{Comparison between RS-L1, RS-CL2, and RS-DL2}
\begin{figure}[hbpt]
\centering
       \includegraphics[width=4in]{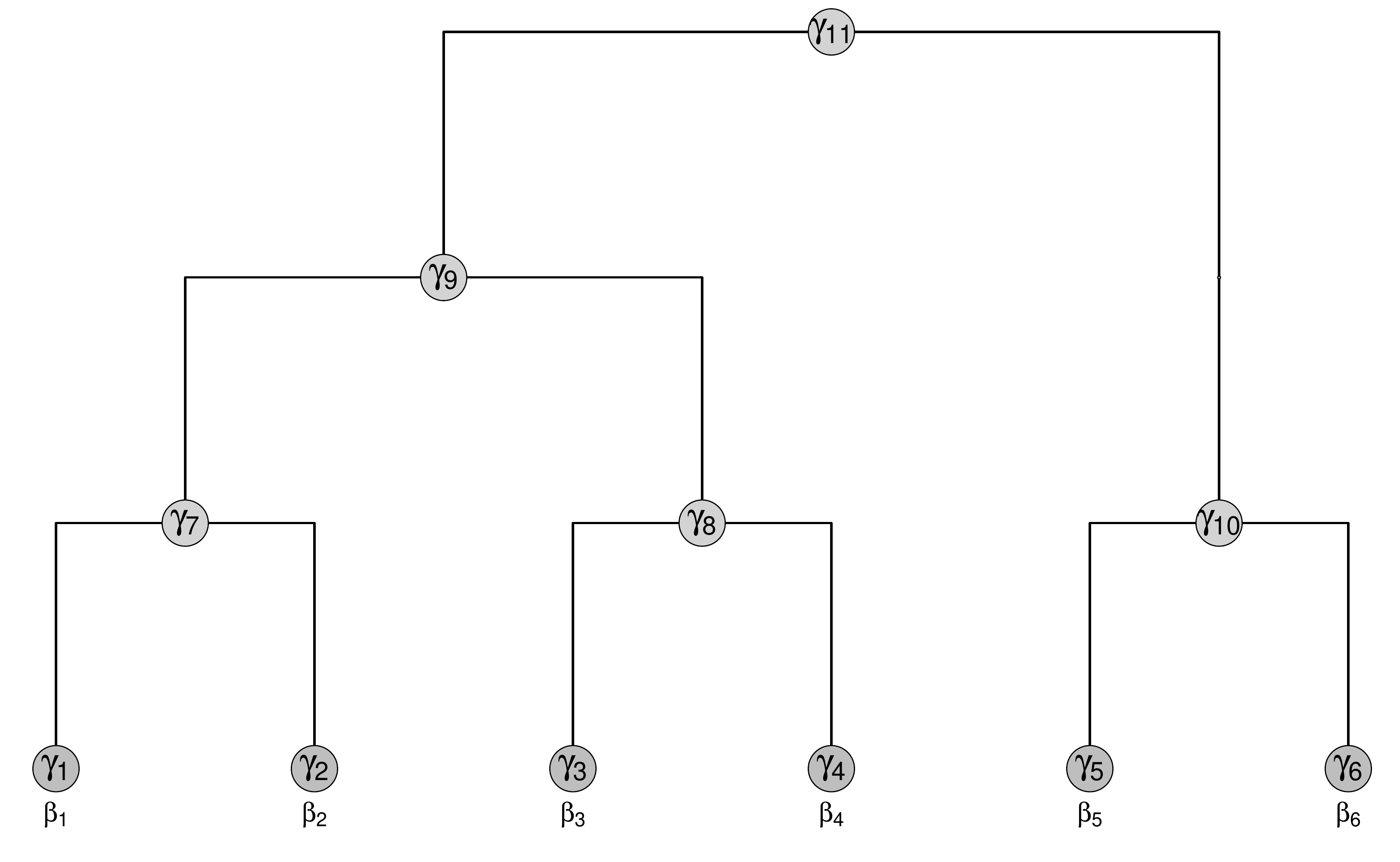}
\vskip-.2in
\caption{The taxonomic tree structure among variables. }
\label{app:fig:tree}
\end{figure}

In this study, we further investigate the differences between RS-L1, RS-CL2, and RS-DL2.
As proof of concept, we focus on a small-scale setting with $p=6$ variables.
The tree structure among the variables is shown in Figure \ref{app:fig:tree}.
We further assume the true coefficients are $\beta_1=\beta_2=\beta_3=\beta_4=0.5$ and $\beta_5=\beta_6=2$.
By definition, if we fix the root node $\gamma_{11}$ to be a constant, only $\gamma_9$ and $\gamma_{10}$ will be nonzero (the actual values of $\gamma_9$ and $\gamma_{10}$ depend on the prefixed value of $\gamma_{11}$).
Following Study I, we simulate the relative abundance data from a logistic Gaussian distribution and generate the response from a relative-shift model (with SNR being 1).
We further truncate the compositional data to get excessive zeros, and use them as input data with measurement errors.

\begin{figure}[hbpt]
\centering
       \includegraphics[width=6in]{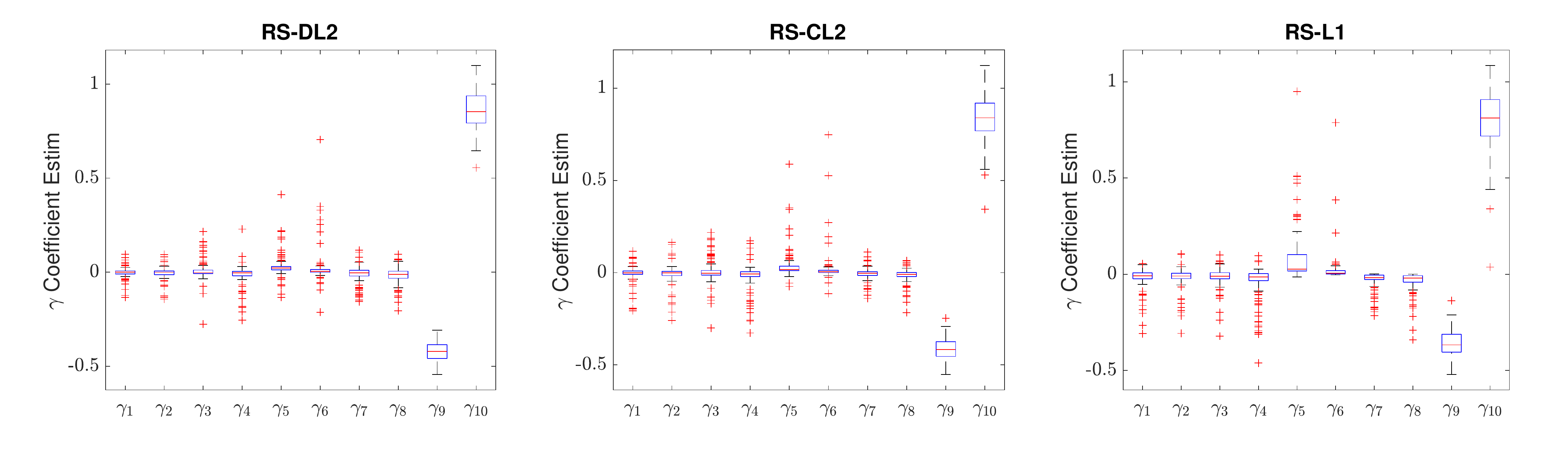}
\vskip-.2in
\caption{The box-plots of intermediate coefficient estimates from RS-DL2, RS-CL2, and RS-L1.}
\label{fig:comp}
\end{figure}

\begin{figure}[hbpt]
\centering
       \includegraphics[width=6.5in]{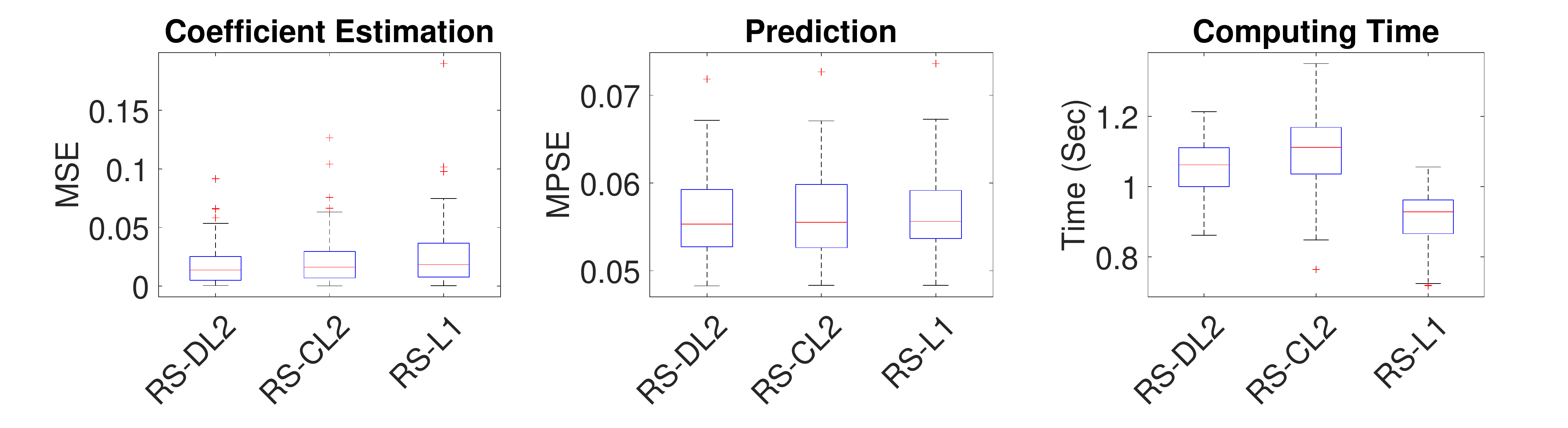}
\vskip-.2in
\caption{The box-plots of $\bbeta$ estimation, prediction, and computing time of RS-DL2, RS-CL2, and RS-L1.}
\label{fig:comp1}
\end{figure}

The comparisons of RS-L1, RS-CL2, and RS-DL2 are shown in Figures \ref{fig:comp} and \ref{fig:comp1}.
In particular, Figure \ref{fig:comp} shows the box-plots of intermediate coefficient estimates from different methods.
Both RS-DL2 and RS-CL2 are slightly better than RS-L1 in obtaining sparse estimation of the intermediate coefficients $\gamma_1$ to $\gamma_8$. This is likely due to the (overlapping) group structure of the penalty terms in {\DL} and {\CL}.
Figure \ref{fig:comp1} further shows the comparison of different methods in $\bbeta$ coefficient estimation, prediction, and computing time.
For $\bbeta$ estimation, RS-DL2 has the smallest mean squared error, followed by RS-CL2.
The difference between each pair of methods is not dramatic, but is statistically significant from a paired t-test at the nominal significance level of 0.05.
The prediction performances of different methods are similar.
The small differences in estimation accuracy may be obscured by the measurement errors in the input data.
For computing time, RS-L1 is the best, likely because this is a rather small-scale setting.
Overall, all three methods provide similar results, and the group penalties in {\DL} and {\CL} may potentially lead to more accurate coefficient estimate.

\section{Proofs of the Theoretical Results}\label{app:proofs}

\begin{proof}[Proof of Theorem 1]

Let $(\widehat \bbeta, \widehat \bgamma)$ be a solution to \eqref{eq:tree-pen}. We have that
\[
\frac{1}{2n} \|\by - \bX \widehat \bbeta\|^2 + \lambda
\mathcal{P}_{T}(\widehat \bgamma) \le \frac{1}{2n} \|\by - \bX \bbeta\|^2 + \lambda
\mathcal{P}_{T}(\bgamma),
\]
for any $(\bbeta, \bgamma)$ such that $\bbeta = \bA
\bgamma$. Recall that $(\bbeta^*, \bgamma^*)$ are the coefficient vectors of the true model satisfying $\bbeta^* = \bA\bgamma^*$. Let $\widehat{\bDelta}^{\gamma} = \widehat{\bgamma} - \bgamma^*$, and $\widehat{\bDelta}^{\beta} = \bA \widehat{\bDelta}^{\gamma} = \widehat{\bbeta} - \bbeta^*$. By using $\by = \bX \bbeta^* + \bvare$, we get
\begin{align}
  \frac{1}{2n}\|\bX \widehat \bbeta - \bX \bbeta^*\|^2
  &\le \lambda \mathcal{P}_{T}(\bgamma^*) - \lambda \mathcal{P}_{T}(\widehat \bgamma) + \frac{1}{n} \bvare^T\bX \widehat{\bDelta}^{\beta}\notag\\
  &= \lambda (\mathcal{P}_T(\bgamma^*) - \mathcal{P}_T(\widehat{\bgamma}) ) + \frac{1}{n}
    \bvare^T\bX \bA {\widehat\bDelta}^{\gamma}.\label{eq:bound1}
\end{align}

We mainly consider the group penalty forms {\CL} and {\DL} in \eqref{pen2} and \eqref{pen3} of the main paper, as the results for the {\NL} penalty in \eqref{pen1} of the main paper can be directly derived from \citet{yan2018rare}. For each node $u \in I(T)$, define $\bP_u$ as a $\real^{(|T| - 1) \times (|T| - 1)}$ diagonal matrix indicating its child nodes, such that the diagonal elements are given by
\[
  (\bP_u)_{vv} =
  \begin{cases}
    1 \quad v \in \mbox{Child(u)}; \\
    0 \quad \mbox{otherwise}.
    \end{cases}
  \]
Similarly, for each node $u \in I(T)$, define $\bM_u$ as a $\real^{(|T| - 1) \times (|T| - 1)}$ diagonal matrix indicating the nodes of its descendants, such that the diagonal elements are given by
\[
  (\bM_u)_{vv} =
  \begin{cases}
    1 \quad v \in \mbox{Descendant(u)}; \\
    0 \quad \mbox{otherwise}.
    \end{cases}
  \]
  Then, the {\CL} penalty in \eqref{pen2} of the main paper can be re-expressed as
  $$
      \mathcal{P}_{T}(\bgamma) =  \sum_{u \in I(T)}\|\bP_u\bgamma\|,
  $$
  and the {\DL} penalty in \eqref{pen3} of the main paper can be re-expressed as
  $$
      \mathcal{P}_{T}(\bgamma) =  \sum_{u \in I(T)}\|\bM_u\bgamma\|.
  $$
  Moreover, the following two identities hold
  \begin{align}
     \sum_{u \in I(T)} \bP_u \bP_u = \bI_{|T|-1}, \qquad \sum_{u \in I(T)} \bP_u \bM_u = \bI_{|T|-1}. \label{eq:pumu}
  \end{align}

With the above treatment of the penalty forms, we are now ready to treat the second term in \eqref{eq:bound1}. Let's focus on the {\CL} penalty first.
\begin{align*}
  |\bvare^T\bX \bA {\widehat\bDelta}^{\gamma}|
  &= |\bvare^T\bX \bA (\sum_{u \in I(T)} \bP_u \bP_u) {\widehat\bDelta}^{\gamma}| \\
  & = |\sum_{u \in I(T)}(\bvare^T\bX \bA\bP_u) (\bP_u{\widehat\bDelta}^{\gamma})| \\
  &\le \sum_{u \in I(T)} \| \bvare^T\bX \bA \bP_u\| \| \bP_u {\widehat\bDelta}^{\gamma}\| \\
  &\le \max_{u \in I(T)} \{\|\bvare^T\bX \bA \bP_u\|\} \sum_{u \in I(T)} \| \bP_u {\widehat\bDelta}^{\gamma}\| \\
  &\le \max_{u \in I(T)} \{\| (\bX \bA \bP_u)^T\bvare\|\}
    \{\sum_{u \in I(T)} \| \bP_u {\widehat\bgamma}\| + \sum_{u \in I(T)} \| \bP_u \bgamma^{*}\|\}.
\end{align*}
That is,
\begin{align}
  |\frac{1}{n}\bvare^T\bX \bA {\widehat\bDelta}^{\gamma}|  \leq \max_{u \in I(T)} \{\| \frac{1}{n}(\bX \bA \bP_u)^T\bvare\|\} (\mathcal{P}_T(\bgamma^*) + \mathcal{P}_T(\widehat{\bgamma}) ).\label{eq:bound3}
\end{align}
It is easy to see that the same inequality holds for the {\DL} penalty due to the second identity in \eqref{eq:pumu}.




Now we bound the stochastic term $\max_{u \in I(T)}\{ \| (\bX \bA \bP_u)^T\bvare\|/n\}$. We will need the following results from Proposition 1 of \citet{hsu2012tail}. 
\begin{lemma}\label{eq:lemma1}
  Let $\bZ$ be an $m \times n$ matrix and let $\Sigma_z = \bZ^T \bZ$. Suppose
  $\bvare = (\epsilon_1,\ldots,\epsilon_n)^T$ is a multivariate Gaussian random with mean zero and covariance $\sigma^2\bI$. For all $t>0$,
  \[
    \Prob(\|\bZ\bvare\|^2 > \sigma^2 \{\mbox{Tr}(\Sigma_z) +
    2\sqrt{\mbox{Tr}(\Sigma_z^2)t} + 2\|\Sigma_z\| t\}) < e^{-t},
  \]
  where $\|\Sigma\|$ denotes the spectral norm of the non-negative definite matrix $\Sigma$.
\end{lemma}

For each $u \in I(T)$, denote $\bX_u = \bX\bA\bP_u \in \mathbb{R}^{n\times p_u}$ with $p_u = |\mbox{Child(u)}|$, and $\Sigma_u = \bX_u\bX_u^T$. It can be recognized that each column of $\bX_u$ is obtained by aggregating the columns of $\bX$ corresponding to the leaf nodes $L(T_{v})$, $v \in \mbox{Child(u)}$. For example, in Figure~\ref{fig:tree} of the main paper, for the internal node $\gamma_{10}$, its $\bX_u$ matrix contains 3 columns, $X_1 + X_2, X_3, X_4$. Since $\bX$ is compositional, it follows that each element of the matrix $\bX_u$ is in the
interval $[0, 1]$, and the sum of each of its rows is less than or equal to 1.
Write $\bX_u =
\{x_{ij}^u\}_{n\times p_u}$, it follows that
\begin{align*}
  &\mbox{Tr}(\Sigma_u) = \sum^n_{i}\sum^{p_u}_{j} (x_{ij}^{u})^2 \le \sum^n_{i}
  (\sum^{p_u}_{j}x_{ij}^{u})^2 \le n; \\
  &\mbox{Tr}(\Sigma_u^2) \le \mbox{Tr}(\Sigma_u)^2 \le n^2;\\
  &\|\Sigma_u\| \le \mbox{Tr}(\Sigma_u) \le n.
\end{align*}

By Lemma \ref{eq:lemma1}, we have that for any $u \in I(T)$ and any $t>0$,
\[
    \Prob\{\|\bX_u^T\bvare\|^2 > n\sigma^2 (1 +
    2\sqrt{t} + 2 t)\} < e^{-t}.
\]
It follows that
\begin{align*}
  \Prob\{\frac{1}{\sqrt{n}}\|\bX_u^T\bvare\| > 2\sqrt{2}\sigma \sqrt{t}\} <
    e^{-t} \mbox{ when } t > 1/2.
\end{align*}
Upon taking $t = \log |I(T)| /\delta > 1 / 2$, we have
\begin{align*}
  \Prob(\frac{1}{\sqrt{n}}\|\bX_u^T \bvare\| > 2\sqrt{2} \sigma\sqrt{\log(|I(T)|)/\delta}) <
    e^{-\log \frac{|I(T)|}{\delta}} = \frac{\delta}{|I(T)|}
\end{align*}
By taking a union bound over all internal nodes, we get
\begin{align*}
  \Prob(\max_{u \in I(T)} \frac{1}{\sqrt{n}}\|\bX_u^T \bvare\| > 2\sqrt{2} \sigma\sqrt{\log(|I(T)|)/\delta} ) \le
    \delta.
\end{align*}
By \eqref{eq:bound1} and \eqref{eq:bound3}, if we take
$\lambda \ge 2 \sqrt{2}\sigma\sqrt{\log(|I(T)|) / (\delta n)}$, then with probability at least $1-\delta$,
\begin{align*}
  \frac{1}{2n}\|\bX \widehat \bbeta - \bX \bbeta^*\|^2
  &\leq \lambda (\mathcal{P}_T(\bgamma^*) - \mathcal{P}_T(\widehat{\bgamma}) ) + \lambda (\mathcal{P}_T(\bgamma^*) + \mathcal{P}_T(\widehat{\bgamma}))\\
  & = 2\lambda \mathcal{P}_T(\bgamma^*).
\end{align*}
The results then follow because the above holds for any $\bgamma^*$ such that $\bbeta^* = \bA\bgamma^*$. This completes the proof.

\end{proof}

\begin{proof}[Proof of Lemma 1]

For the {\NL} penalty in \eqref{pen1} of the main paper, the result directly follows based on the construction of the coarsest aggregating set $B^*$ and a corresponding choice of $\bgamma^*$, i.e., for each $u \in B^*$, $\gamma_u^* = \beta_j^*$ for any $j \in L(T_u)$, and otherwise $\gamma_u^* = 0$. 
It is easily seen that this $\bgamma^*$ satisfies $\bA\bgamma^* = \bbeta^*$. It then follows that $\sum_{u\in T_{-r}}|\gamma_u| \leq  M|B^*|$.




For the {\CL} penalty, consider again the above construction of $\bgamma^*$ according to the coarsest aggregating set $B^*$. Immediately, we have that
  $$
  \|\bgamma^*\| \leq M\sqrt{|B^*|}.
  $$
Define
  $$
  B^*_p = \{u\in I(T); \mbox{Child(u)}\cap B^* \neq \emptyset.\}.
  $$
  That is, $B^*_p$ collects the parent nodes of all the nodes in $B^*$. We have that $\|(\gamma_v)_{v\in \mbox{\mbox{Child(u)}}}\| \neq 0$ only if $u \in B_p^*$. Since there is no overlap among $\mbox{Child(u)}$, $u \in B_p^*$, we also have that $|B^*_p| \leq |B^*|$. It then follows that
\begin{align*}
  \sum_{u \in I(T)}\|(\gamma_v)_{v\in \mbox{\mbox{Child(u)}}}\|
  =
    \sum_{u \in B_p^*}\|(\gamma_v)_{v\in \mbox{\mbox{Child(u)}}}\|
   \leq \sqrt{|B_p^*|}\|\bgamma^*\|
   \leq M\sqrt{|B_p^*||B^*|} \leq M|B^*|.
\end{align*}
Here we use the fact that for any partition of a vector, i.e., $\ba= (\ba_1^T,\ldots, \ba_s^T)^T$, it holds that $\sum_{i=1}^{s}\|\ba_i\|\leq \sqrt{s}\|\ba\|$. 
This completes the proof.

\end{proof}

\begin{proof}[Proof of Corollary 1]
Due to Assumption 3, we have that $|I(T)| = O(p)$. The results then following immediately by combining the results in Lemma 1 and Theorem 1.

\end{proof}

\bibliographystyle{chicago}
\bibliography{biblio_composition}

\end{document}

%% file: Notation.tex
\newcommand{\real}{\mathbb{R}} 
\newcommand{\simp}{\mathbb{S}} 
\newcommand{\pen}{\mathcal{P}} 

\DeclareMathOperator*{\argmin}{arg\,min}

\def\ba{\boldsymbol{a}}

\def\bc{\boldsymbol{c}}

\def\bx{\boldsymbol{x}}
\def\by{\boldsymbol{y}}

\def\bA{\boldsymbol{A}}

\def\bC{\boldsymbol{C}}
\def\bD{\boldsymbol{D}}

\def\bI{\boldsymbol{I}}

\def\bM{\boldsymbol{M}}

\def\bP{\boldsymbol{P}}

\def\bX{\boldsymbol{X}}

\def\bZ{\boldsymbol{Z}}

\newcommand{\1}{\mathbf{1}}

\newcommand{\balpha}{\boldsymbol{\alpha}}
\newcommand{\bbeta}{\boldsymbol{\beta}}

\newcommand{\bgamma}{\boldsymbol{\gamma}}

\newcommand{\bvare}{\boldsymbol{\varepsilon}}
\newcommand{\bxi}{\boldsymbol{\xi}}

\newcommand{\bDelta}{\boldsymbol{\Delta}}

\newcommand{\be}{\begin{equation}}
\newcommand{\ee}{\end{equation}}
\newcommand{\bea}{\begin{eqnarray}}
\newcommand{\eea}{\end{eqnarray}}
\newcommand{\bes}{\begin{eqnarray*}}
\newcommand{\ees}{\end{eqnarray*}}
